\documentclass{aa}
\usepackage{graphicx}
\usepackage{txfonts}
\usepackage{longtable,lscape,deluxetable_mod}
\usepackage{natbib}
\bibpunct{(}{)}{;}{a}{}{,} 
\newcommand{\asca}{{\emph{ASCA}}}
\newcommand{\rosat}{{\emph{ROSAT}}}

\newcommand{\sdss}{{\emph{SDSS}}}
\newcommand{\spitzer}{{\emph{Spitzer}}}

\newcommand{\xmm}{{XMM-\emph{Newton}}}
\newcommand{\hb}{\ensuremath{\mbox{H}\beta}}
\newcommand{\mgii}{\ensuremath{\mbox{\ion{Mg}{ii}}}}
\newcommand{\oiii}{\ensuremath{\mbox{[\ion{O}{iii}]}}}
\newcommand{\apix}[2]{\ensuremath{#1^{\,\mbox{\scriptsize #2}}}}
\newcommand{\pedix}[2]{\ensuremath{#1_{\,\mbox{\scriptsize #2}}}}
\newcommand{\pedap}[3]{\ensuremath{#1_{\,\mbox{\scriptsize #2}}^{\,\mbox{\scriptsize #3}}}}
\newcommand{\ie}{i.e.}
\newcommand{\eg}{e.g.}
\newcommand{\csam}{``control sample''}
\newcommand{\msam}{``main sample''}
\newcommand{\kev}{\ensuremath{\,\mbox{\scriptsize keV}}}
\newcommand{\ang}{\ensuremath{\,\mbox{\scriptsize \AA}}}
\newcommand{\nh}{\ensuremath{\mbox{cm}^{-2}}}
\newcommand{\nhsym}{\ensuremath{N_{\mbox{\scriptsize H}}}}
\newcommand{\flux}{\ensuremath{\mbox{ergs~cm}^{-2}\mbox{~s}^{-1}}}
\newcommand{\fluxA}{\ensuremath{\mbox{ergs~cm}^{-2}\mbox{~s}^{-1}\mbox{~\AA}^{-1}}}
\newcommand{\lum}{\ensuremath{\mbox{ergs~s}^{-1}}}
\newcommand{\lumA}{\ensuremath{\mbox{ergs~s}^{-1}\mbox{~\AA}^{-1}}}
\newcommand{\lumHz}{\ensuremath{\mbox{ergs~s}^{-1}\mbox{~Hz}^{-1}}}
\newcommand{\mum}{\ensuremath{\,\mu\mbox{\scriptsize m}}}
\begin{document}
   \title{Exploring X-ray and radio emission of type~1 AGN up to $z \sim 2.3$\thanks{Tables
    3--5 are only available in electronic form at the CDS via
    anonymous ftp to  cdsarc.u-strasbg.fr (130.79.128.5)
    or via http://cdsweb.u-strasbg.fr/cgi-bin/qcat?J/A+A/}}


   \author{
          L. Ballo\inst{\ref{inst1}}
          \and
          F.J.H. Heras\inst{\ref{inst1}}
	  \fnmsep\thanks{Present Address: Department of Zoology, University of Cambridge, Downing St, Cambridge, CB2 3EJ, UK}
          \and
          X. Barcons\inst{\ref{inst1}}
          \and
          F.J. Carrera\inst{\ref{inst1}}
          }

  \offprints{L.~Ballo, \email{ballo@ifca.unican.es}}

   \institute{Instituto de F{\'{\i}}sica de Cantabria (CSIC-UC), Avda. Los Castros s/n (Edif. Juan Jord\'a), 
   E-39005 Santander  (Spain)\label{inst1}
             }

   \date{Received XXXXXXXXXX XX, XXXX; accepted XXXXXXXXXX XX, XXXX}

  \abstract
   {X-ray emission from Active Galactic Nuclei (AGN) is dominated by the accretion disk around a supermassive black
   hole. 
   The radio luminosity, however, has not such a clear origin except in the most powerful sources where jets are 
   evident. 
   The origin (and even the very existence) of the local bi-modal distribution in radio-loudness is also a debated 
   issue.}
   {By analysing X-ray, optical and radio properties of a large sample of type 1 AGN and quasars (QSOs) up to $z> 2$,
   where the bulk of this population resides, we aim to explore the interplay between radio and X-ray emission in AGN, 
   in order to further our knowledge on the origin of radio emission, and its relation to accretion.}
   {We analyse a large ($\sim 800$ sources) sample of type 1 AGN and QSOs selected from the 2XMMi \xmm\ X-ray
   source catalogue, cross-correlated with the \sdss\ DR7 spectroscopic catalogue, covering a redshift range from $z\sim
   0.3$ to $z\sim 2.3$. Supermassive black hole masses are estimated from the \mgii\ emission line, bolometric
   luminosities from the X-ray data, and radio emission or upper limits from the FIRST catalogue.}
   {Most of the sources accrete close to the Eddington limit and the distribution in radio-loudness does not appear to 
   have a bi-modal behaviour.
   We confirm that radio-loud AGN are also X-ray loud, with an X-ray--to--optical ratio up to twice
   that of radio-quiet objects, even excluding the most extreme strongly jetted sources.  
   By analysing complementary radio-selected control
   samples, we find evidence that these conclusions are not an effect of the X-ray selection, but are
   likely a property of the dominant QSO population.}
   {Our findings are best interpreted in a context where radio emission in AGN, with the exception of a minority of
   beamed sources, arises from very close to the accretion disk and is therefore heavily linked to X-ray emission. We
   also speculate that the radio-loud/radio-quiet dichotomy might either be an evolutionary effect that developed well
   after the QSO peak epoch, or an effect of incompleteness in small samples.}


   \titlerunning{}
   \authorrunning{L. Ballo et al.}

   \maketitle
%

\section{Introduction}\label{sect:intro}

Strong X-ray and radio emissions are properties that distinguish active galactic nuclei (AGN) from the whole population
of galaxies.
X-rays are the most direct manifestation of the accretion disk around a supermassive black hole (SMBH) at the centre of
the galaxy hosting the AGN.
Although radio emission is most apparent in a fraction of AGN, in particular in those classified as ``radio-loud'' (RL),
which constitute $10-20$\% of the local AGN population, recent work shows that even radio-quiet (RQ) AGN
exhibit a radio-emitting core, which might result from some sort of radio plasma arising in the vicinity of the SMBH.
While optical emission in AGN is 
due to the superposition of thermal components coming from different distances from the nucleus, with a contribution of radiation reprocessed 
outside the AGN central engine,
both X-rays
and radio emission can be used to probe the immediate environment of the SMBH.

From the observational point of view, two quantities appear most relevant in exploring the possible link between radio
emission and accretion: the UV-based radio-loudness, defined as 
$\mathcal{R}\equiv\pedix{F}{5\,GHz,\,rf}/\pedix{F}{2500\,\AA,\,rf}$ 
\citep[monochromatic fluxes in the rest frame;][]{stocke92}, 
and the Eddington ratio $\pedix{\lambda}{Edd}\equiv\pedix{L}{bol}/\pedix{L}{Edd}$, where $\pedix{L}{bol}$ is the 
bolometric luminosity and
$\pedix{L}{Edd}\equiv1.3\times 10^{38}\,\pedix{M}{BH}/\pedix{M}{\sun}\,$[\lum] is the limiting luminosity of Eddington
\citep{eddington13,rees84}.
Obtaining these quantities requires radio, UV and X-ray fluxes, plus reliable bolometric corrections 
and SMBH mass (\pedix{M}{BH}) estimates.
In contradiction to earlier works, \citet{ho02} showed that $\mathcal{R}$ is uncorrelated with \pedix{M}{BH}, but
strongly
anticorrelated  with $\pedix{\lambda}{Edd}$.
This was interpreted by \citet{ho02} in the framework of changes of the accretion mode, from a radiatively efficient
standard accretion at $\pedix{\lambda}{Edd}> 0.01$ to a radiatively inefficient ADAF (Advection Dominated Accretion 
Flow, which is prone to radio-emission) at lower Eddington ratios.

The local low-luminosity AGN population appears to show a dichotomy in radio-loudness, where two distinct populations 
appear to peak at $\mathcal{R}\sim 10-100$ (RL) and $\mathcal{R}\sim 0.1-1$ \citep[RQ;][]{kellerman89}.
This bimodality is not so apparent (or plainly non-existing) in deeper radio \citep{white00} or X-ray selected surveys
\citep{brinkmann00}, where AGN samples display a continuous distribution in radio-loudness.
It is then unclear whether there is something fundamentally different between strongly and weakly emitting radio AGN,
and this is directly linked to the origin of radio emission in these objects.

With mass accretion rate largely regulating the luminosity of the AGN, the only other relevant parameter to modulate
radio-loudness and whether an AGN develops or not a powerful jet, is the SMBH spin \citep{blandford90,wilson95}.
In a simple scenario where jets are formed in rapidly spinning SMBH, evolution
by galaxy and associated SMBH mergers
would naturally lead towards a SMBH population with low spin \citep{berti08}.
This would imply that smooth accretion (which tends to spin up SMBH) would be unimportant in the history of SMBH growth.
However, the dependence of RL fraction on redshift and luminosity is still a strongly debated issue \citep[\eg,][and references therein]
{jiang07}: the number of RL sources  in the analyzed samples, not large enough to study
their two-dimensional distribution in redshift and luminosity, and the wide range of selection criteria used to define the samples observed 
contribute to a large range of contradicting results.

\citet{sikora07} extend this view by showing that the anticorrelation between $\mathcal{R}$ and $\pedix{\lambda}{Edd}$ 
comes in two parallel tracks, one for RL AGN residing in elliptical galaxies and one (lower) for RQ AGN residing in
spiral galaxies.
They propose a revised spin paradigm, where elliptical galaxies (and thence RL AGN) host highly spinning SMBH as
a result of at least one major merger in the past, while spiral galaxies (and thence RQ AGN) underwent mostly
chaotic accretion, \ie, accretion of small mass fragments with random angular momenta.
But since highly accreting luminous QSOs residing in ellipticals are largely RQ, speculations that radio
emission might be intermittent have been put forward.
Most recent versions of the spin paradigm call for retrograde systems (where the SMBH and the accretion disk counter
rotate) as the mechanism to extract the most powerful jets \citep{garofalo10}.
RL AGN are mostly assumed to be retrograde, and RQ prograde.
Natural evolution tends to make all SMBH-accretion disk systems prograde, which would explain the overwhelmingly large
fraction of RQ AGN in the local Universe.

The release of large catalogues of fairly deep X-ray and radio sources, along with the optical photometry and
spectroscopy provided by the Sloan Digital Sky Surveys 
\citep[\sdss; see][for the final public data release from \sdss-II]{abazajian09}, 
calls now for a study of the relation between accretion and radio properties in large samples of AGN.
This is of particular interest to infer the physical origin of the radio emission.
In this study we use the incremental Second \xmm\ Serendipitous Source Catalogue \citep[2XMMi;][]{watson09},
which we correlate with the \sdss\ Data Release~7 (DR7) to select a sample of X-ray detected type~1 AGN and QSOs.
In this way, we use independent data to estimate the bolometric luminosity (from the X-ray data and suitable bolometric
corrections) and the SMBH mass that we estimate from the \sdss\ spectra using the \mgii\ broad emission line.
The latter effectively limits our sample to $z\leq 2.3$, which is however enough to encompass the peak of the QSO
activity epoch at around $z\sim 2$.
Note that we do not include type~2 AGN in this study, in part because the mass estimates based on other proxies (\eg,
the \oiii\ line width) are more uncertain and would compromise the homogeneity of our study.
Radio information is obtained from the FIRST-VLA catalogue \citep{first}.

The paper is organized as follows.
In Section~\ref{sect:sample} we describe the selection method used to generate the samples; Section~\ref{sect:nucl} is 
devoted to recover the nuclear properties (\ie, SMBH masses, nuclear bolometric luminosities, and Eddington ratios).
In Section~\ref{sect:rl} radio-loudness properties are discussed and compared with
results from the literature. 
Finally, in Section~\ref{sect:concl} we summarize our work.
Throughout this paper, a
concordance cosmology with $\pedix{H}{0}=71\,$km s$^{-1}$ Mpc$^{-1}$,
$\pedix{\Omega}{\ensuremath{\Lambda}}=0.73$, and 
$\pedix{\Omega}{m}=0.27$ \citep{spergel03,spergel07} is adopted. 
The energy spectral index, $\alpha$, is defined such that
$\pedix{F}{\ensuremath{\nu}} \propto \apix{\nu}{-\ensuremath{\alpha}}$. 
The photon index, defined as $\pedix{N}{\ensuremath{\epsilon}} \propto
\apix{\epsilon}{-\ensuremath{\Gamma}}$, where $\epsilon$ is the photon energy,
is $\Gamma=\alpha+1$.


\section{Sample selection}\label{sect:sample}

The starting point for the present work is the cross-correlation of the 2XMMi with the \sdss\ DR7 carried out by 
\citet{pineau10}.

The 2XMMi catalogue is an extended version of the 2XMM catalogue, covering $\sim 1$\% of the sky \citep{watson09}; in 
their analysis, \citet{pineau10} considered only point-like sources having a positional error smaller or equal to 
$5\arcsec$, obtaining an initial sample of $\sim 200000$ unique 2XMMi sources.

To construct the initial list of optical candidate counterparts, among the more than $350$ million distinct objects 
contained in the \sdss\ DR7 Photometric Catalog \citep{abazajian09}, only the so-called primary sources were considered.
To perform the correlation, a query centred in the centre of the FOV of each \xmm\ observation was then run, with a 
search radius equal to the distance from the centre to the farthest X-ray source increased by $3\arcmin$.

The counterpart identification was performed by computing a likelihood ratio, defined as the probability of finding the 
optical counterpart at a normalized distance $r$ divided by the probability of having a spurious object at the same 
distance.
The applied formalism, aimed at providing probabilities of identification based on positional coincidences only (no 
other information such as spectral energy distribution were used) led to $30055$ X-ray sources with more than $90$\% 
probability of identification in the DR7.
At this threshold, \citet{pineau10} estimated only $2$\% spurious matches and a $77$\% completeness.

We were interested in weighting the black hole in the centre of a sample of AGN using the width of optical lines emitted 
from the Broad Line Region (BLR), namely the \mgii$\,\lambda 2799\,$\AA\ (see Sect.~\ref{sect:nucl}).
Among the X-ray selected sources detected in the \sdss, 
we then considered objects that were targets of optical spectroscopic 
follow-up in the \sdss, and classified as {\sc qso} or {\sc galaxy}\footnote{We do not consider {\sc hiz\_qso} 
({\tt specClass}$\,=4$), whose high redshift ($z>2.3$) implies the exit of the \mgii\ line from the \sdss\ spectrum.} 
({\tt specClass} parameter equal $3$ or $2$, 
respectively\footnote{See http://cas.sdss.org/astrodr7/en/help/docs/enum.asp?n=SpecClass}).
Among those, we limited ourselves to the $906$ showing in their spectra a \mgii\
line broad enough to have its origin in the 
BLR, specifically FWHM(\mgii)$\,>900\,$km~s$^{-1}$.

Recently, \citet{shen10} presented a compilation of properties of the sources in the \sdss\ DR7 quasar catalogue 
\citep{schneider10}; $892$ out of the $906$ sources just mentioned are in this catalogue, the rest being Seyfert
galaxies.
The fit to the \mgii\ line in $28$ of those $892$ provides null values for the FWHM or the 
\pedix{M}{BH}: we then removed these sources from our sample.
We are therefore left with $878$ X-ray selected type~1 AGN ($864$ QSOs and $14$ with Seyfert-like luminosity).

We collected information about radio emission for our sample by cross-correlating it with the 
Faint Images of the Radio Sky at Twenty-cm \citep[FIRST;][]{first} survey.
The sky coverage of the FIRST survey implies that only $837$ sources ($823$ QSOs) out of $878$ fall in the
FIRST fields, with $100$ detections ($98$ among the QSOs).
Being interested in studying the radiation arising from their nuclear regions, only the core emission should be
considered.
Following \citet{shen10}, $27$ out of $98$ radio-detected QSOs have multiple FIRST source matches within $30$\arcsec.
Comparing the contribution to the radio emission at the \sdss\ source position expected from the different FIRST
components, we found that for all but one of these QSOs with multiple matched FIRST sources, the off-nuclear contamination is not
negligible ($\gtrsim 1$\%): these $26$ not core-dominated QSOs were therefore removed.
A visual inspection of the FIRST images confirms that the rest of the detected QSOs and the $2$ radio-detected sources 
with Seyfert-like luminosity have only one FIRST source within the FIRST resolution.

Our final sample (hereafter, \msam) is composed by $852$ X-ray selected type~1 AGN ($838$ QSOs and $14$ with
Seyfert-like luminosity); out of them, $811$ fall in the FIRST fields, with $74$ detections ($72$ among the QSOs).
A summary of this sample is presented in Table~\ref{tab:sum}.

%
\begin{table}
\begin{minipage}[!ht]{\columnwidth}
\caption{Summary of the samples.} \label{tab:sum}             
\begin{center}          
\renewcommand{\footnoterule}{}  
{\scriptsize
\begin{tabular}{c@{\extracolsep{-0.05cm}}  c  c | c | c}
\vspace{0.2cm}\\
 \multicolumn{3}{l|}{}& \msam & \csam \\
 \hline  
 \hline  
  \multicolumn{3}{c|}{TOTAL}                                  & $852$                            & $4508$ \\
 \hline  
  \multicolumn{3}{c|}{in FIRST}                               & $811$                            & $4508$ \\
  \multicolumn{3}{c|}{(radio det.\quad -\quad radio undet.)}  & ($74$\quad -\quad $737$)         & ($4508$\quad -\quad $0$) \\
  \multicolumn{1}{l}{Radio class.:} & \multicolumn{2}{l|}{RL} & $59$                             & $3796$ \\
  & \multicolumn{1}{r|}{} & \multicolumn{1}{l|}{det. RI}      & $15$                             & $702$ \\
  & \multicolumn{1}{r|}{non-RL} & \multicolumn{1}{l|}{$1 <\mathcal{R}^{up.lim}\leq 10$}  & $443$ & $0$ \\
  & \multicolumn{1}{r|}{} & \multicolumn{1}{l|}{RQ}	      & $7$                              & $10$ \\
  & \multicolumn{2}{l|}{{\it n.c.}}		              & $287$                            & $0$ \\
 \hline  
  \multicolumn{3}{c|}{in \rosat}                              & $69$                             & $501$ \\
  \multicolumn{3}{c|}{in \rosat\ \& in FIRST}                 & $67$                             & $501$ \\
  \multicolumn{3}{c|}{(radio det.\quad -\quad radio undet.)}  & ($13$\quad -\quad $54$)          & ($501$\quad -\quad $0$) \\
  \multicolumn{1}{r}{Radio class.:} & \multicolumn{2}{l|}{RL} & $9$                              & $398$ \\
  & \multicolumn{1}{r|}{} & \multicolumn{1}{l|}{det. RI}      & $4$                              & $99$ \\
  & \multicolumn{1}{r|}{non-RL} & \multicolumn{1}{l|}{$1 <\mathcal{R}^{up.lim}\leq 10$}  & $35$  & $0$ \\
  & \multicolumn{1}{r|}{} & \multicolumn{1}{l|}{RQ}	      & $5$                              & $4$ \\
  & \multicolumn{2}{l|}{{\it n.c.}}		              & $14$                             & $0$ \\
\end{tabular}
}
\end{center}          
\end{minipage}
\end{table}

\begin{figure}
 \centering
 \resizebox{\hsize}{!}{\includegraphics{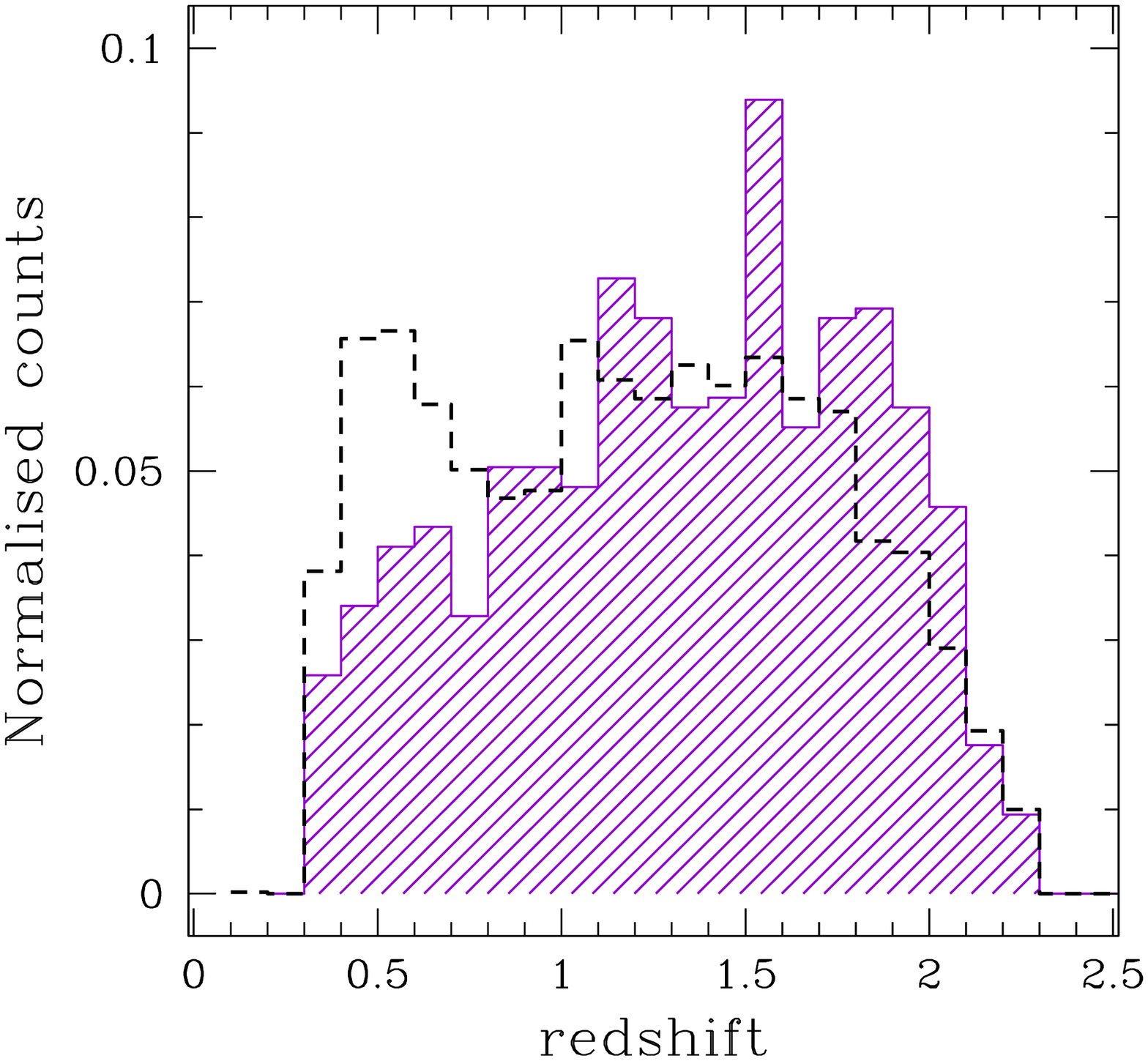}\includegraphics{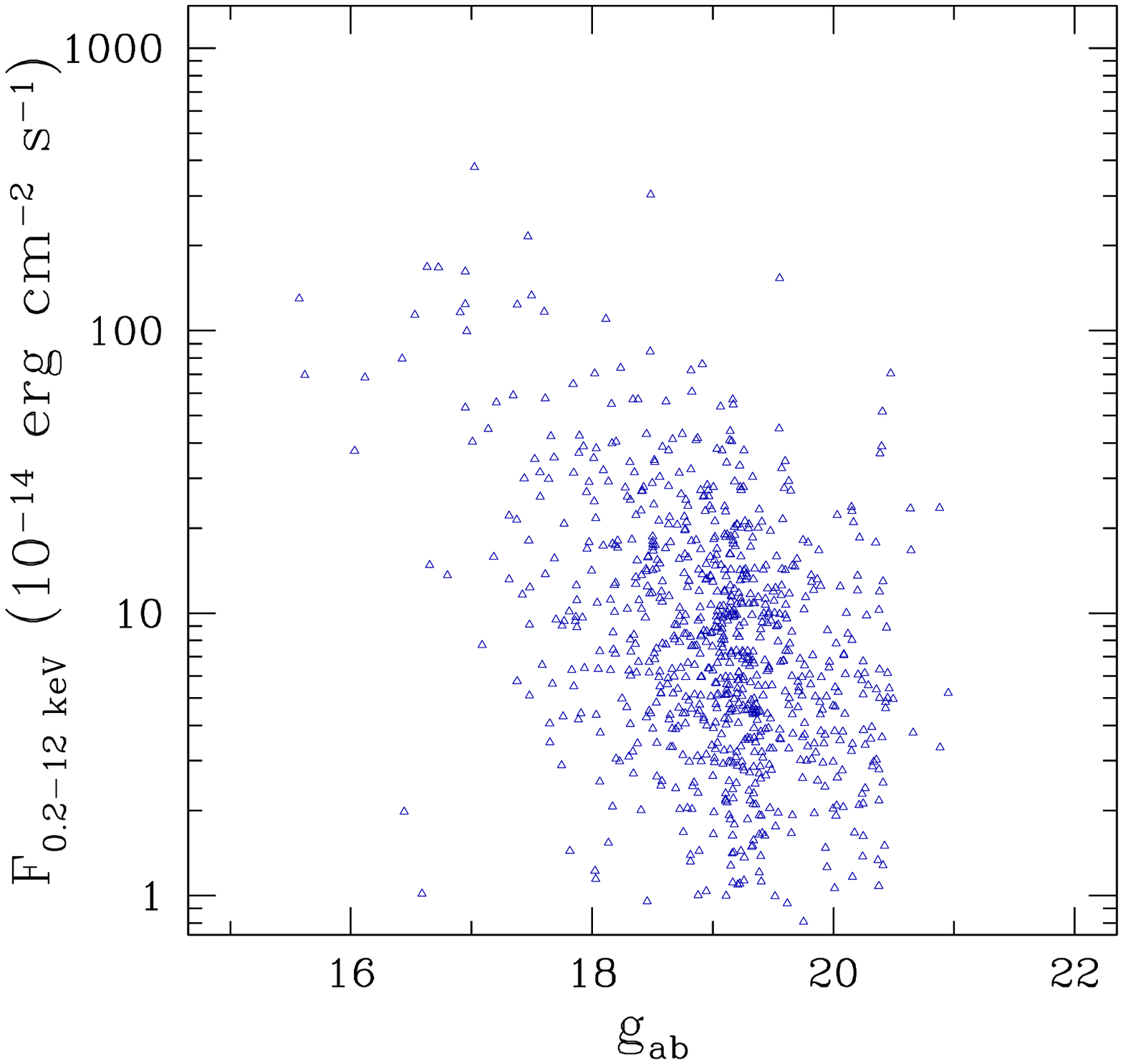}}
 \caption{Properties of the X-ray selected type~1 AGN in our sample. 
 {\it Left panel}: 
 Normalized distribution in redshift (magenta shaded histogram); the black dashed line shows the
 distribution for the \csam.
 {\it Right panel}: Total X-ray fluxes vs. $g$-band AB magnitudes.}
 \label{fig:sample}%
\end{figure}

Basic information is listed in Table~\ref{tab:sample}, only available in electronic form at the CDS\footnote{See
http://cdsweb.u-strasbg.fr/A+A.htx}; for convenience of the reader, we show here a portion.
As noted before, the requirement of covering with the \sdss\ spectra an energy range containing the \mgii\ line imply a
cut in distance; our sources have redshifts more or less uniformly distributed between $0.3$ and $2.3$ 
(see Fig.~\ref{fig:sample}, {\it left panel}).
In Figure~\ref{fig:sample} ({\it right panel}) the sample distribution in the optical--X-ray flux plane is reported.

We note that the main characteristic of our sample is to be X-ray-selected. 
In order to check for possible selection biases, and to test our findings in terms
of nuclear properties, we constructed a \csam\ by searching for
\sdss\ radio sources matching the same optical constraints adopted to construct our main sample, either falling in a
region not covered by \xmm, or not detected in X-rays.
We found $5888$ objects in the \sdss-FIRST cross-correlation having \mgii\ line widths and optical spectroscopic
classifications satisfying the criteria previously described, not included in the 2XMMi catalogue; $5548$ of them are
classified as QSOs by \citet{schneider10}. 
Rejecting the $188$ QSOs for which \citet{shen10} do not provide valid black hole masses from the \mgii\ line, 
we are left with $5700$ sources.
In the same way as we have done for the \msam, we calculated the off-nuclear contamination for the $1196$ sources showing multiple matches in
the FIRST images within $30$\arcsec; this analysis drove us to remove from the sample all but four not core-dominated 
AGN.
The \csam\ is therefore composed by $4508$ radio-detected type~1 AGN ($4204$ QSOs and $304$ with
Seyfert-like luminosity); their redshift distribution is overplotted in Figure~\ref{fig:sample} ({\it left panel}),
while
their classification in terms of radio properties is summarized in Table~\ref{tab:sum}.

Only $11$ out of this \csam\ ($7$ classified as QSOs) fall in one or more of the \xmm\ fields
used to construct the 2XMMi, and
show positive ``good exposure time'' in at least one of the EPIC cameras.
Thanks to {\sc FLIX}\footnote{See http://www.ledas.ac.uk/flix/flix.html}, the flux upper limit server for \xmm\ data
provided by the \xmm\ Survey Science Center (SSC), we obtained for 
all these X-ray undetected AGN
an upper limit in the $0.2-12\,$keV flux.
Basic information for these $11$ sources is reported in the second part of Table~\ref{tab:sample}.

From this point of view, useful information can be obtained considering the match between the two samples and the 
\rosat\  All-Sky Survey \citep[RASS;][]{rass}.
In the \msam, $13$ out of the $69$ \rosat-detected sources are radio-detected; $501$ objects in the \csam\ are
\rosat-detected.

%
%


\section{Recovering the nuclear properties}\label{sect:nucl}

The spectral information provided by the 2XMMi and the \sdss~DR7 catalogues has been used to study the nuclear activity
of the galaxies in our sample.
The black hole (BH) masses were derived from the optical measurements (continuum and line width), while we estimated 
the bolometric luminosity from the X-ray flux.
The use of data from different energy ranges allows an independent determination of the two parameters.
A discussion of the error estimates is presented in  Sect.~\ref{sect:err}.
Optical- and radio-based parameters have been calculated also for the \csam; X-ray-based quantities have 
been obtained for the $11$ sources for which upper limits to the X-ray flux were obtained.

%
\begin{figure*}
 \centering
 \resizebox{\hsize}{!}{\includegraphics{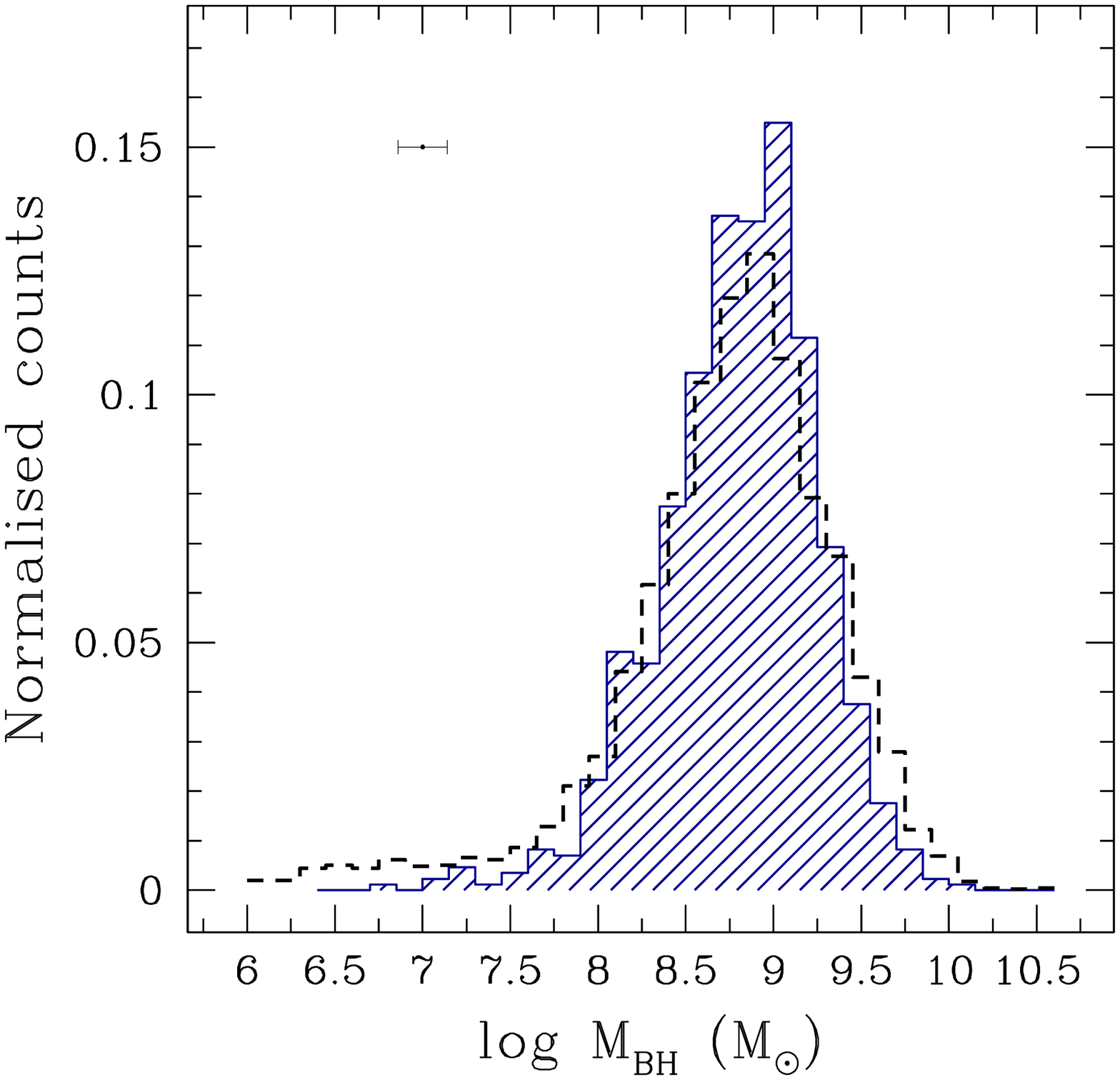}
 \includegraphics{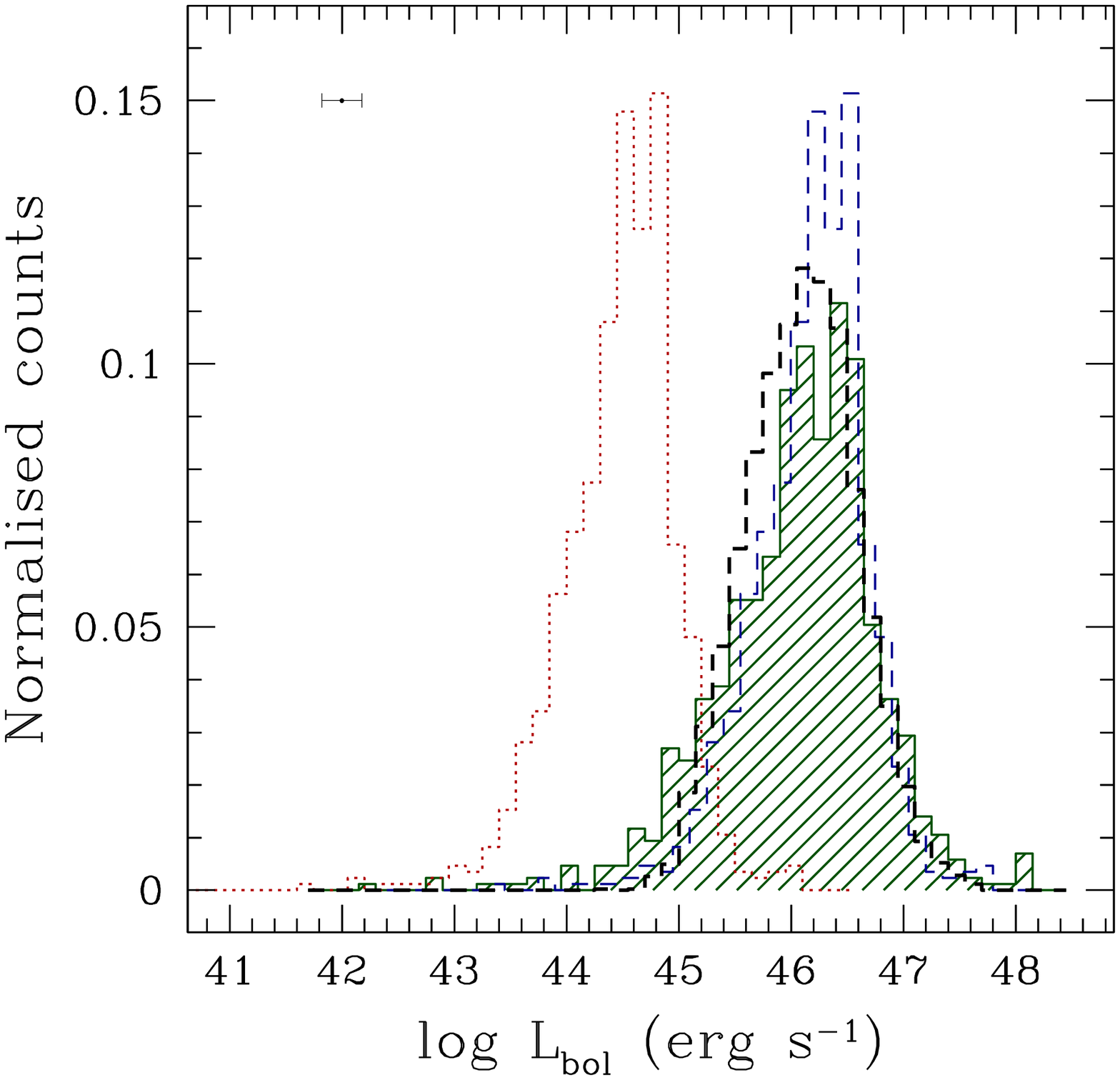}
 \includegraphics{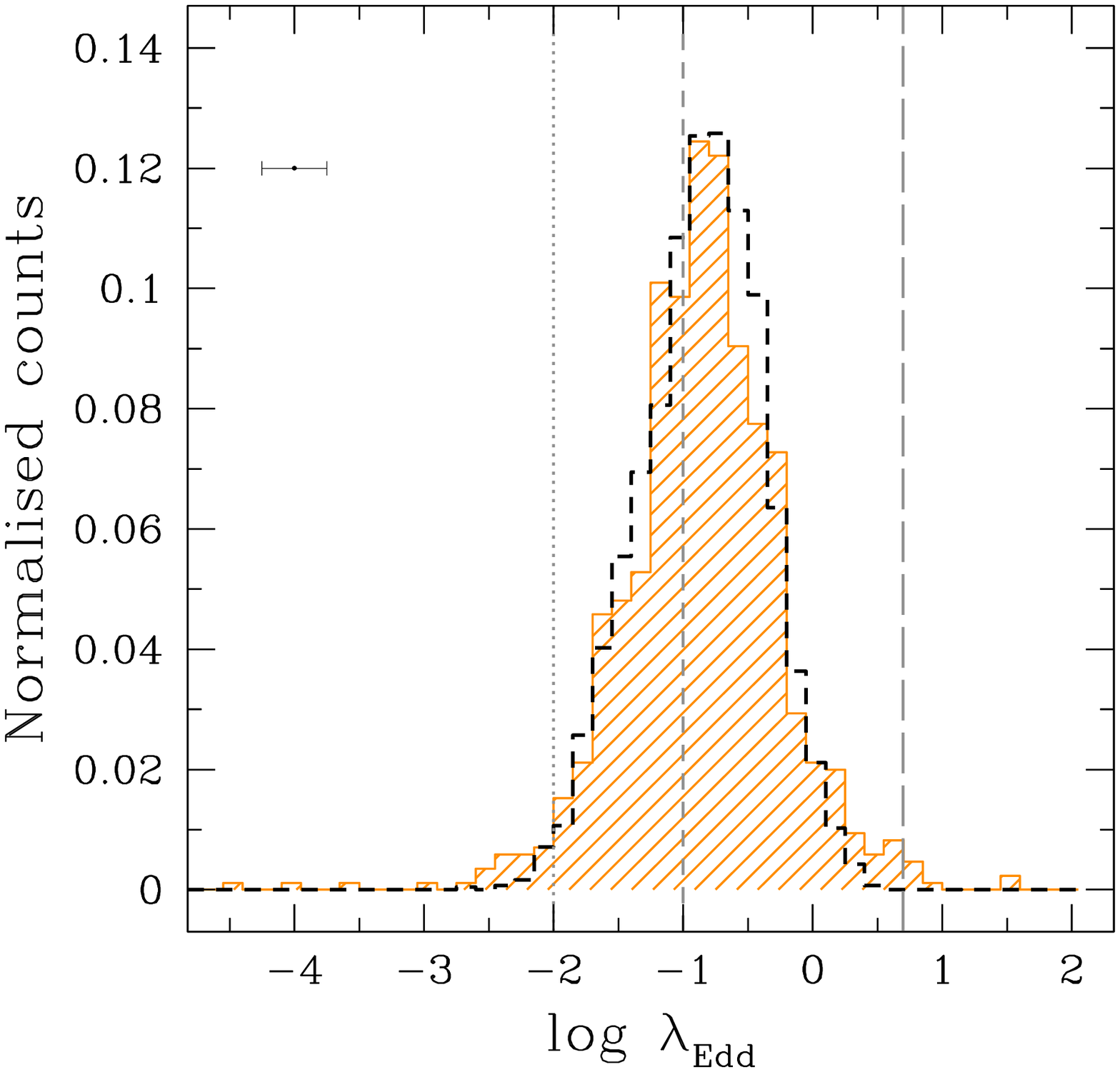}}
 \caption{Nuclear properties, derived as described in Sect.~\ref{sect:nucl}; the error bars reported represent the mean
 uncertainties obtained as detailed in Sect.~\ref{sect:err}.
{\it Left panel}: Normalized distribution in BH mass (blue shaded histogram); the black dashed line shows the
 distribution for the \csam.
{\it Central panel}: Distribution in bolometric luminosity obtained assuming the luminosity-dependent
 \pedix{\kappa}{2-10\kev} in equation~(\ref{eq:lbol}), green shaded histogram; the red dotted line and the blue dashed line represent
 the distributions in bolometric luminosity obtained assumed a constant $\pedix{\kappa}{2-10\kev}=1$ (\ie, the distribution of the intrinsic 
 hard X-ray luminosity) and $\pedix{\kappa}{2-10\kev}=50$, respectively.
 The black dashed line shows the distribution in bolometric luminosity for the \csam, computed from optical/UV luminosities \citep[][]{shen10}.
 {\it Right panel}: Distribution in Eddington ratio (yellow shaded histogram);  the black dashed line shows the distribution in Eddington ratio 
 for the \csam\
 \citep[][]{shen10}. Grey dotted, dashed, and long-dashed vertical lines define the 
 regions of $\pedix{\lambda}{Edd} < 0.01$, $0.1 < \pedix{\lambda}{Edd} < 5$, and high-super-Eddington accretion, 
 respectively.}
 \label{fig:derdist}%
\end{figure*}

\subsection{Black hole masses}\label{sect:mbh}

For each source, the mass of the central compact object was estimated from the so-called mass-scaling relations 
using broad-emission-line widths and nuclear continuum luminosities 
\citep[and references therein]{kaspi00,kaspi05,vestergaard06,vestergaard08}.
Being interested in exploring in a uniform way (\ie, using the same proxy for all the sources) the BH masses and
accretion rates in a sample of AGN extended up to (relatively) high redshift, the best choice is the relation based on
the \mgii\ emission line.
For the subsample of QSOs, we adopted the mass obtained by 
\citet[{\tt LOGBH\_MGII\_VO09} column in their catalogue]{shen10} adopting the recalibration proposed by 
\citet{vestergaard09}:
   \begin{equation}\label{eq:mbh}
      \pedix{M}{BH}=10^{6.86}\left[\frac{FWHM(\mgii)}{1000\,\mbox{km~s}^{-1}}\right]^2
      \left[\frac{\lambda\pedix{L}{$\lambda$}}{10^{44}\,\lum}\right]^{0.5}
   \end{equation}
for $\lambda=3000\,$\AA, with a $1\sigma$ scatter in the logarithmic zero-point of $0.55\,$dex.

For the $14$ X-ray selected AGN (and for the $304$ sources in the \csam) not present in the QSO catalogue, 
we adopted the same relation; the intrinsic FWHM was obtained from the observed standard deviation of the \mgii\ line, 
as reported in the \sdss\ catalogue, while 
the monochromatic continuum luminosity was estimated using a typical QSO template.
We adopted the composite spectrum provided by P.~Francis\footnote{See
http://msowww.anu.edu.au/pfrancis/composite/widecomp.d} in 2002, updating the original template presented by
\citet{francis91}.
To normalize the template, we used the continuum flux observed under the line, as reported in the \sdss\ catalogue (see
Table~\ref{tab:sample}).
The derived line widths, continuum luminosities, and BH masses are reported in Table~\ref{tab:deriv} (also in this case,
the entire table is available in electronic form).
We found rather high masses, narrowly distributed between $\sim 10^8$ and $\sim 3\times10^9\,$\pedix{M}{\sun} (see
Fig.~\ref{fig:derdist}, {\it left panel}; the distribution for the \csam\ is overplotted).

\subsection{Bolometric luminosities}\label{sect:lbol}

Concerning the bolometric luminosity, 
several bolometric corrections, starting from the emission at various wavelengths, can be found in the literature.
The results obtained from such relations must be taken with care, being based on average SEDs, assumed to describe the broad-band 
emission of quite different sources; moreover, the emission observed in each energy range could be affected by contamination due to 
different external and/or reprocessed components, and these contributions could be extremely difficult to evaluate.
Nevertheless, they provide a powerful tool to investigate the physics of the nucleus.
Here, we took advantage of having high-energy information, directly linked to the
AGN innermost regions and less affected by obscuration.
We derived the hard X-ray luminosity in the $2-10\,$keV energy range from the EPIC-pn $0.2-12\,$keV flux (provided in the 2XMMi catalogue) corrected for
the Galactic absorption, assuming as intrinsic emission a power law with $\Gamma$ fixed to $1.9$.

Note that in case of absorption, the intrinsic emission would be more intense than estimated here; a similar result
would be produced if the real observed spectrum is flatter than $1.9$.
Therefore, the X-ray luminosity adopted here can be in principle lower than the intrinsic one 
\citep[see \eg\ ][]{panessa06}.
From the analysis of the hardness-ratios presented in Sect.~\ref{sect:eddrrl}, we expect a mean correction factor
$\pedap{L}{2-10\kev}{abs.PL}/\pedap{L}{2-10\kev}{PL}$ of
$\sim 1.1$, with a maximum value of $\sim 2.8$.

To estimate the bolometric luminosity, we assumed the luminosity-dependent X-ray bolometric correction
($\pedix{\kappa}{2-10\kev}\equiv\pedix{L}{bol}/\pedix{L}{2-10\kev}$) following \citet{marconi04}: 
   \begin{eqnarray}\label{eq:lbol}
      \log \left(\pedix{L}{bol}/\pedix{L}{2-10\kev}\right)&=1.54 + 0.24\left(\log \pedix{L}{bol}-45.58\right) \\
      &+ 0.012\left(\log \pedix{L}{bol}-45.58\right)^{2} \nonumber\\
      &- 0.0015\left(\log \pedix{L}{bol}-45.58\right)^{3} \nonumber
   \end{eqnarray}
with a $1\sigma$ scatter $\sim0.1$ (taken by the authors to be independent of the luminosity).

In Figure~\ref{fig:derdist} ({\it central panel}) we present the distribution of bolometric luminosities (green shaded
histogram; the red dotted line superimposed represents the distribution of intrinsic X-ray luminosity in the
$2-10\,$keV energy range.
The former is clearly broader, as expected due to the assumption of an increasing luminosity-dependent bolometric
correction.
The effects of possible intrinsic absorption on the bolometric luminosities are slightly higher than in the case of the
X-ray luminosity, although still rather low (mean correction factor $\pedap{L}{bol}{abs.PL}/\pedap{L}{bol}{PL}\sim 1.2$, 
with a maximum value of $\sim 4.2$).

Our results critically depend on the bolometric luminosities recovered, then in turn on the correction applied to the
X-ray luminosities.
The selection of the best X-ray bolometric correction is not a trivial issue.
Depending on the studied sample, the range of the electromagnetic spectrum used to reconstruct the SEDs, and whether the
reprocessed emission was considered, different solutions have been proposed.
Some authors considered luminosity-dependent \pedix{\kappa}{2-10\kev}, \citep[\eg, ][]{marconi04,hopkins07},
suggesting changes in the physics of the disk-corona system with the intensity of the nuclear emission,
whereas others found corrections with a very shallow or absent correlation with X-ray luminosity
\citep[\eg, ][]{elvis94,richards06,marchese09}.
As expected due to the luminosity-dependent expression adopted, the \pedix{\kappa}{2-10\kev} distribution shows a significant spread 
(ranging from $\sim5$ up to few hundred).
However, although slightly broadened at low values, our distribution of \pedix{L}{bol} from the luminosity-dependent correction 
from \citet{marconi04} is roughly consistent with that obtained assuming a mean constant value 
$\pedix{\kappa}{2-10\kev}=50$ (see Fig.~\ref{fig:derdist}, {\it central panel}, blue dashed line).

\begin{figure}
 \centering
 \resizebox{\hsize}{!}{\includegraphics{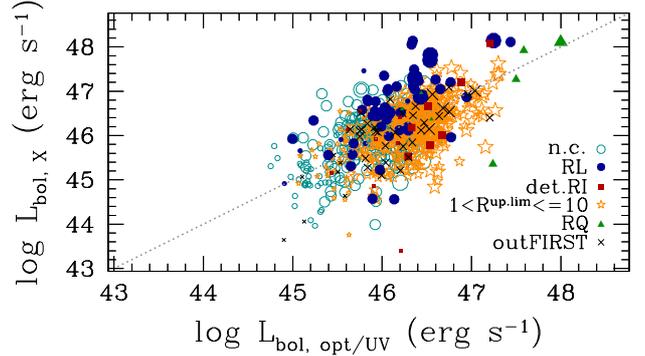}}
 \caption{ \pedix{L}{bol} inferred from X-ray luminosity vs. \pedix{L}{bol} derived from the optical/UV continuum
   for the sub-sample of QSOs \citep{shen10}.
   Different symbols mark radio classification (see Sect.~\ref{sect:rl}): not classified, light-blue open circles; 
   RL, blue filled circles; detected RI, red filled squares; undetected sources with $\mathcal{R}^{up.lim}\in (1,10]$,
   yellow open stars; RQ, green filled triangles; out of FIRST, black crosses.
   The symbol size increases with the redshift ($z<0.7$, $0.7\leq z<1.9$, and $z\geq 1.9$), therefore identifying the
   continuum wavelength adopted to obtain the bolometric luminosity (see the text).
   The dotted line marks the one-to-one relation between the luminosities.}
 \label{fig:chklbol}%
\end{figure}

Although their origin in the innermost region of the AGN makes the X-rays the most direct proxy of the total emission, 
we have reworked our calculations with the bolometric luminosities obtained from optical/UV observations.
We considered the values reported by \citet{shen10}, computed from optical/UV continuum luminosities at different
wavelength depending on the source redshift, $\pedix{L}{5100\AA}$ at $z<0.7$, $\pedix{L}{3000\AA}$ at $0.7\leq z<1.9$,
and $\pedix{L}{1350\AA}$ at $z\geq 1.9$, and adopting the bolometric corrections from the composite SED in
\citet{richards06}.
As noted by the authors, the global SED from \citet{richards06} also counts the infrared (IR) bump in estimating the
bolometric corrections; removing the IR radiation, assumed to come from the reprocessed UV radiation, reduces
the bolometric corrections by about one third.
In the same way, we estimated a bolometric luminosity for the QSOs in the \csam, which lack X-ray information; its distribution is overplotted 
as black dashed line in Figure~\ref{fig:derdist} ({\it central panel}).
In Figure~\ref{fig:chklbol}, for the QSOs in the \msam\ we compare our X-ray based $\pedix{L}{bol}$ with that derived from
the optical/UV, with the correction for the IR contribution applied.
We note a slight departure from the one-to-one relation for sources with important radio emission with respect to the
optical/UV radiation (RL; see below) in the highest redshift bin ($z\geq 1.9$).
Nevertheless, the good agreement found for these objects among the two estimates of \pedix{L}{bol} (consistent at a
$3\sigma$ level) over a significant range in luminosity reassures us on the reliability of 
our approach.

%

\subsection{Eddington ratios}\label{sect:eddr}

The analysis performed allowed us to investigate the accretion powering these systems.
In particular, Eddington ratios $\pedix{\lambda}{Edd}$ can be calculated as the ratio between bolometric and Eddington 
luminosities, 
where $\pedix{L}{Edd}\equiv1.3\times 10^{38}\,$\pedix{M}{BH}/\pedix{M}{\sun}~[\lum] represents the exact balance 
between inward gravitational force and outward radiation force acting on the gas.
Since the AGN luminosity is directly proportional to the accretion rate, $\pedix{\lambda}{Edd}$ is a measure of the 
accretion rate relative to the critical Eddington value.

\begin{figure}
 \centering
 \resizebox{\hsize}{!}{\includegraphics{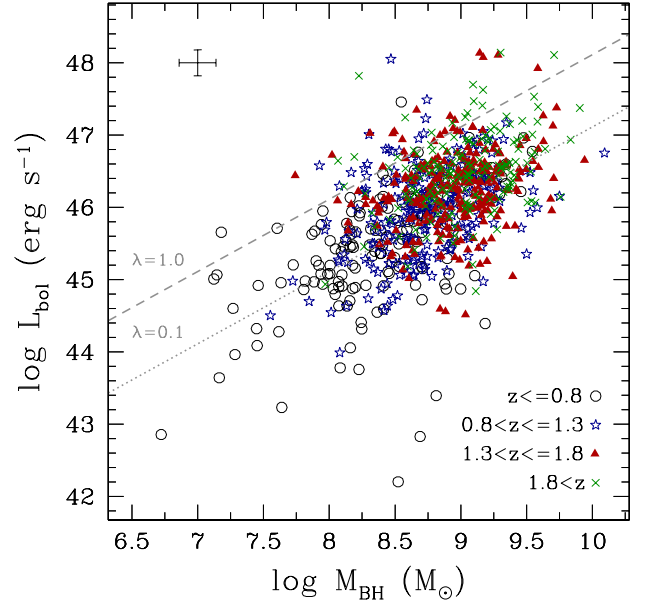}}
 \caption{Bolometric luminosities vs. black hole masses for the sample of X-ray emitting type~1 AGN, divided in 
 intervals 
 of redshift ($z \le 0.8$, black open circles; $0.8 < z \le 1.3$, blue open stars; $1.3 < z \le 1.8$, red filled 
 triangles; $z>1.8$, green crosses). 
 Only mean error bars are reported to avoid clutter.
 Grey dashed and dotted lines define the locus for sources emitting at the Eddington limit and at $1/10$ of it.
}
 \label{fig:lbolmbh}%
\end{figure}

We found rather high Eddington ratios (see Fig.~\ref{fig:derdist}, {\it right panel}, and last column in
Table~\ref{tab:deriv}): for more than half of the \msam\ we found $\pedix{\lambda}{Edd}$ between $0.1$ and $1$, while 
only $3$\% show
$\pedix{\lambda}{Edd} \leq 0.01$.
If we were to correct X-ray luminosities for the intrinsic absorption estimated from the hardness-ratio analysis, would only
increase $\pedix{\lambda}{Edd}$.
No significant trend with the redshift is observed (see Fig.~\ref{fig:lbolmbh}).

Super-Eddington accretion is found for $54$ sources.
Luminosities exceeding the Eddington limit can be observed if accretion is not spherically symmetric
\citep{osterbrock89}; it has been suggested that accretion disks with radiation-driven inhomogeneities could produce
luminosities exceeding the Eddington value \citep{begelman02}.
Eddington ratios close to, or even higher than, unity are often estimated for the subclass of Narrow-line Seyfert 1
galaxies \citep[\eg,][]{collin02,collin04}.

\subsection{Error budget}\label{sect:err}

Among the measurements obtained from the different catalogues used in this work, error estimates are available for the
FWHM(\mgii), the X-ray flux in the $0.2-12\,$keV energy range observed frame, and the radio flux at $1.5\,$GHz observed
frame.
Moreover, in our determination of the nuclear parameters, we made a number of assumptions that must be taken into 
account when discussing the global uncertainties.\\

Errors for the \pedix{M}{BH} 
are provided by \citet[{\tt LOGBH\_MGII\_VO09\_ERR} column in their catalogue]{shen10}; for the sources not comprised in
their catalogue (in the \msam\ and in the \csam, $14$ and $304$ objets, respectively), the errors
are evaluated considering three different contributions\footnote{We note that no errors on
the flux under the line are provided in the \sdss\ tables.}:
\vspace{-0.2cm}
\begin{itemize}
  \item The uncertainties associated with the width of the \mgii\ line, provided by the \sdss\ catalogue, typically of 
  the 
  order of $\leq 40$\%.
  \item A systematic error induced by the selection of one spectral
shape common to all the sources for the optical emission: for
each source, starting from a spectral shape $\pedix{F}{$\lambda$} \propto \apix{\lambda}{$\beta$}$ we determine
the scatter in the optical luminosity due to a different choice in
the slope ($\beta=-1.37\pm0.25$, averaging out the rest-frame spectra of $215$ \sdss\ QSO; F.~Fontanot, priv. comm.).
The scatter in the \pedix{L}{$\lambda$} (@$3000\,$\AA), calculated assuming as a pivot the flux below the \mgii\ line,
is of $\sim 1.7$\%.
  \item The scatter in the phenomenological relation described by equation~(\ref{eq:mbh}).
\end{itemize}
\vspace{-0.2cm}
When propagated to yield the uncertainties of the BH masses, the third factor always prevails over the first two. 
Considering the whole \msam, the uncertainties on $\pedix{M}{BH}$ are lower than $37$\%
\\

Uncertainties in the optical fluxes at $2500\,$\AA\ and $4400\,$\AA\ rest-frame (see Sect.~\ref{sect:rl}) were obtained
assuming again $\Delta\beta=\pm0.25$ for the slope in the optical continuum, with fluxes below the \mgii\ 
and \hb$\,\lambda 4861\,$\AA\ lines as pivot, respectively.
The resulting scatters are $\sim 2.8$\% and $\sim 2.5$\%, respectively.\\

We identify two possible sources of uncertainty in our estimate of the bolometric luminosity:
\vspace{-0.2cm}
\begin{itemize}
  \item The error in the normalization in equation~(\ref{eq:lbol}).
  \item The uncertainties in the X-ray luminosities, obtained from the error on the flux provided by the 2XMMi
  catalogue.
This contribution affects the bolometric luminosities in a non-linear way, given the luminosity-dependent bolometric
correction.
\end{itemize}
\vspace{-0.2cm}
The resulting uncertainties on $\log \pedix{L}{bol}$ are lower than $5$\%; through quadratical sum propagation, we
obtain a mean error on $\log \pedix{\lambda}{Edd}$ of $\sim 0.5$.\\

Finally, for the radio flux we need to combine two sources of error:
\vspace{-0.2cm}
\begin{itemize}
  \item The uncertainties associated with the FIRST radio flux.
  \item A systematic error induced by the selection of one spectral index common to all the sources (see
  Sect.~\ref{sect:rl}), estimated assuming a change in the radio spectral index of $\sim 25$\%.
\end{itemize}
\vspace{-0.2cm}
The relative errors, $\pedix{\sigma}{\pedix{F}{5\,GHz}}/\pedix{F}{5\,GHz}$, are lower than $28$\%  for both the 
$74$ radio-detected sources and the $737$ radio-undetected sources, for which we assumed the flux limit of the FIRST 
survey (see Sect.~\ref{sect:rl}), respectively.
This translates in the following  mean relative errors for the radio-loudness and the X-ray loudness parameters 
(defined as $\pedix{\mathcal{R}}{X}\equiv\pedix{{\nu}L}{5\,GHz,\,rf}/\pedix{L}{2-10\kev,\,rf}$; see
Sect.~\ref{sect:sed}):
$\pedix{\sigma}{\ensuremath{\mathcal{R}}}/\ensuremath{\mathcal{R}}$ and 
$\pedix{\sigma}{\pedix{\ensuremath{\mathcal{R}}}{X}}/\pedix{\ensuremath{\mathcal{R}}}{X}$, of $0.11$
and $0.3$ (radio-detected), and $0.10$ and $0.19$ (radio-undetected), respectively.
 
%
%


\section{Radio loudness}\label{sect:rl}

%
\begin{figure*}
 \centering
 \resizebox{\hsize}{!}{\includegraphics{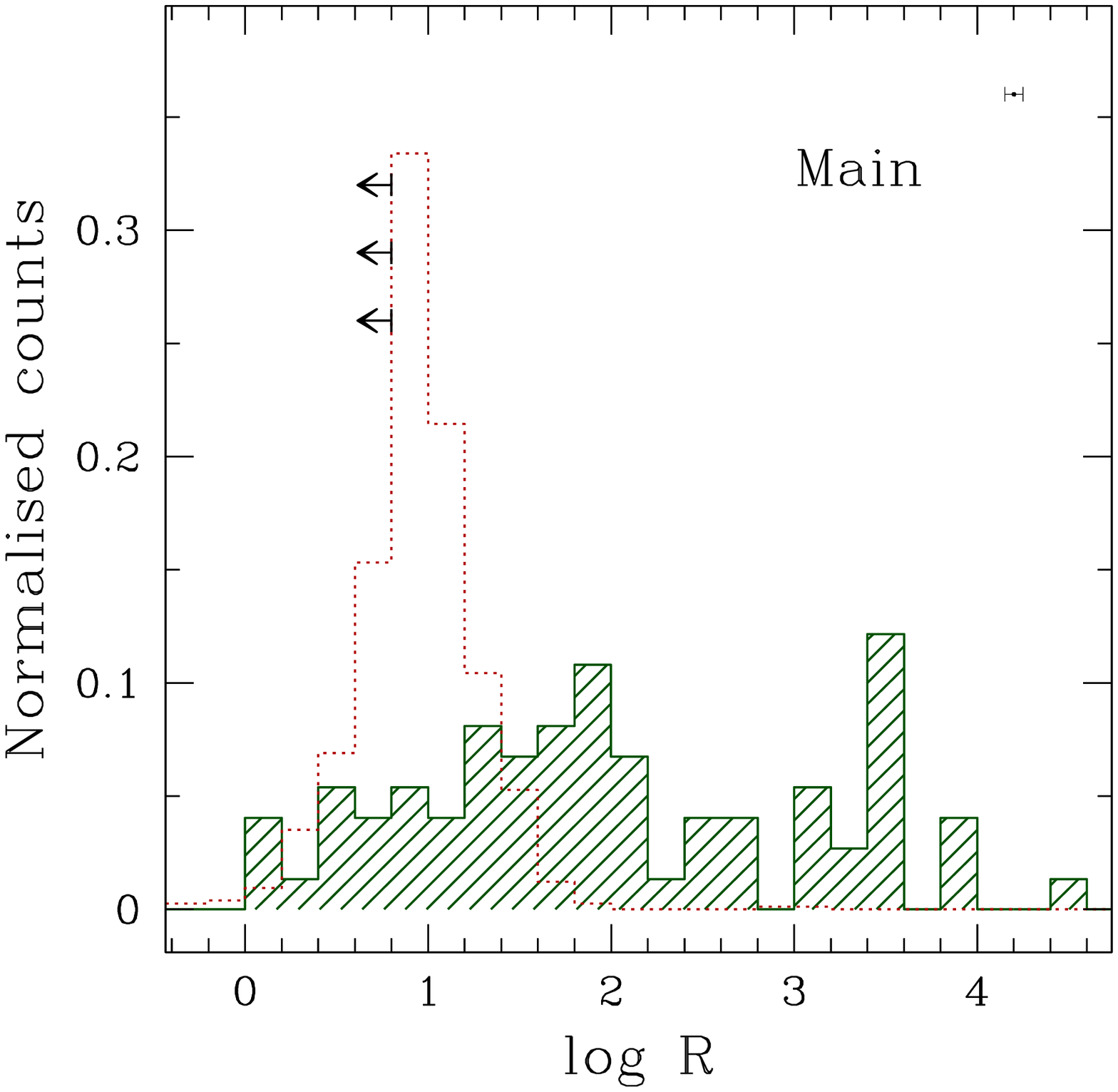}
 \includegraphics{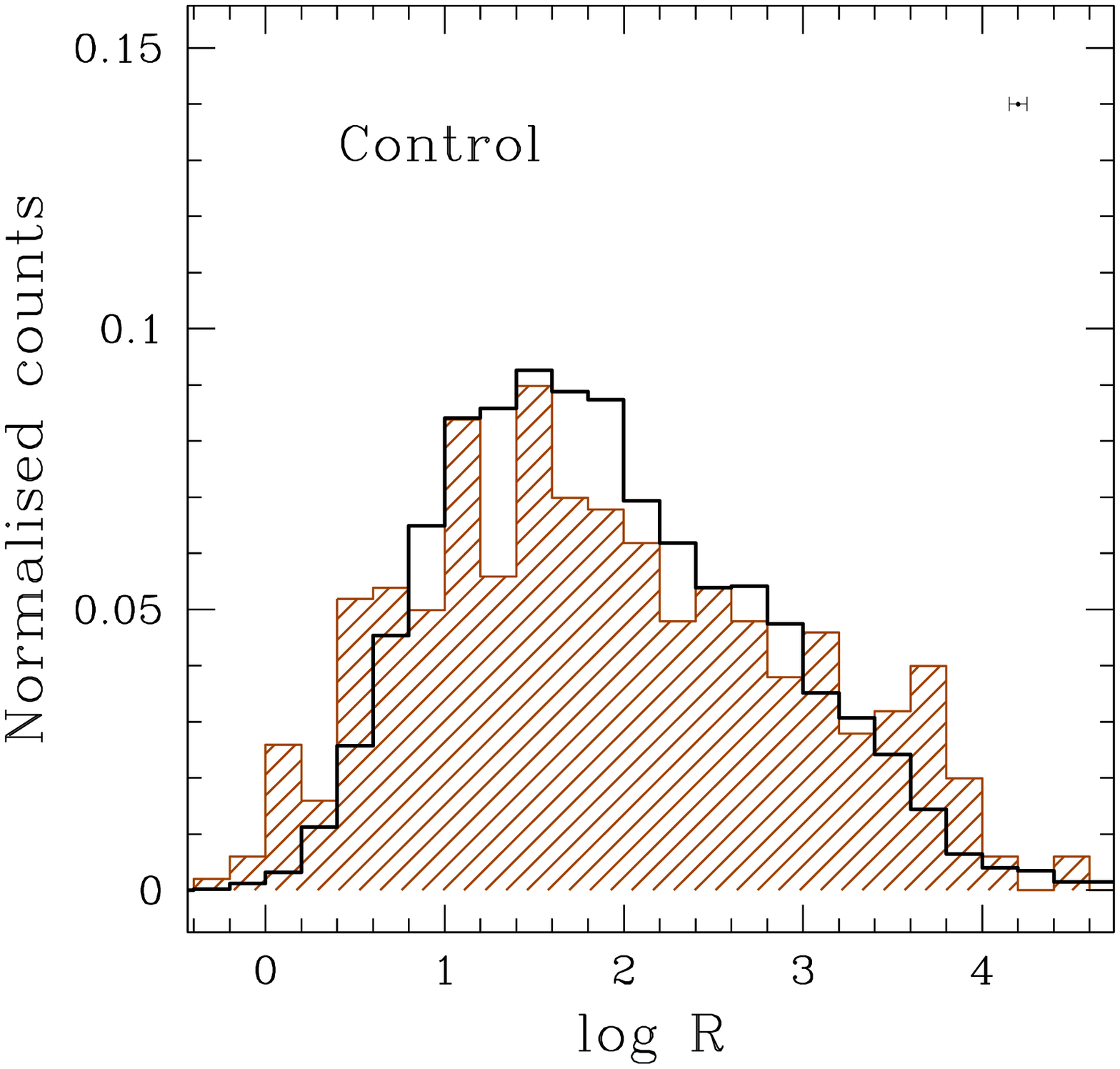}
 \includegraphics{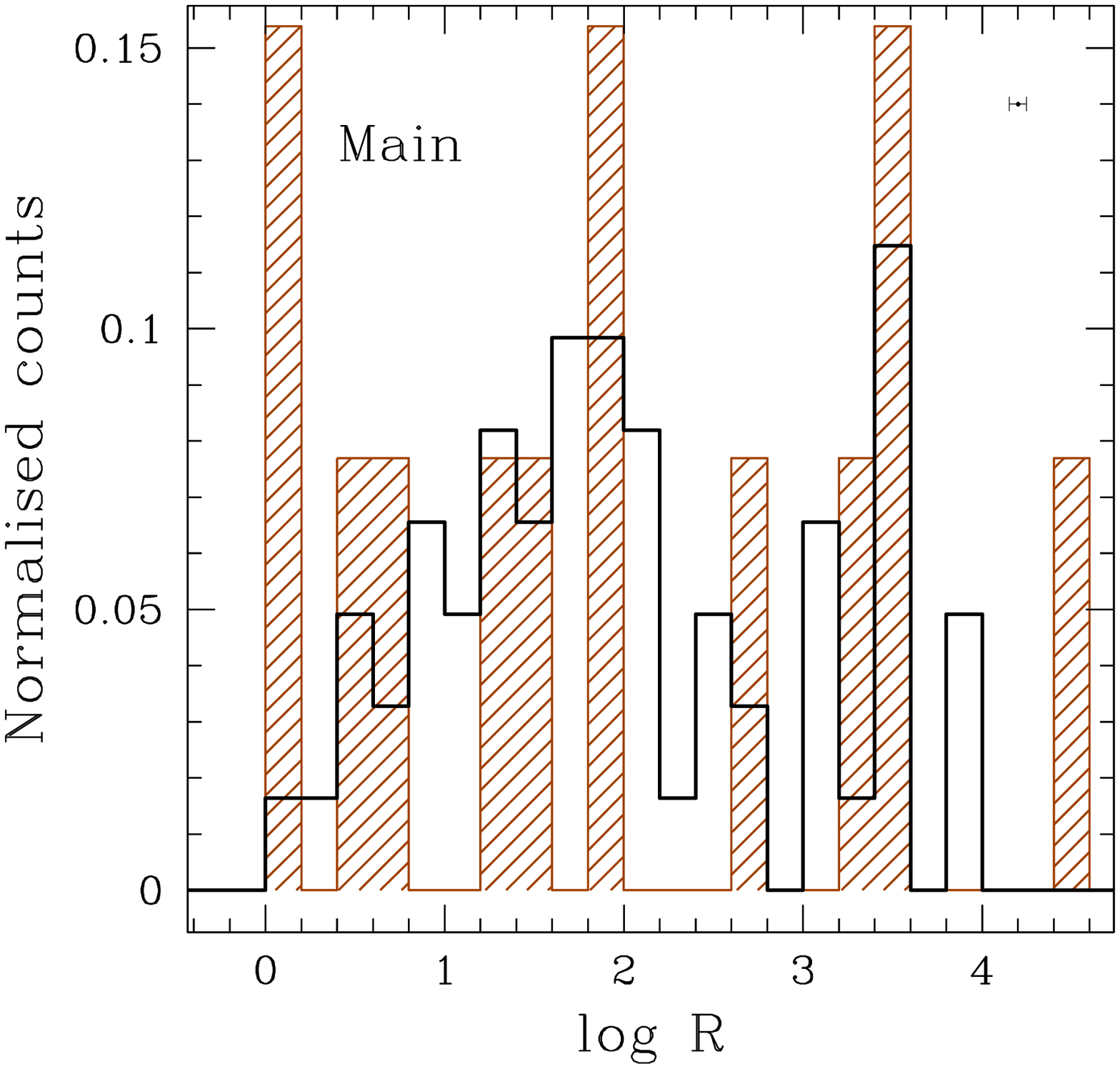}}
 \caption{Normalised distributions of radio-loudness parameter $\mathcal{R}$, derived as described in 
 Sect.~\ref{sect:nucl}; the error bars reported represent the mean uncertainties obtained as detailed in
 Sect.~\ref{sect:err}.
 {\it Left panel}: radio-detected (green shaded histogram) and radio-undetected (red dotted line) sources in the \msam. 
 {\it Central panel}: \rosat-detected (brown shaded histogram) and \rosat-undetected (black continuous line) sources in 
 the \csam.
 {\it Right panel}: radio-detected sources in the \msam, divided between \rosat-detected (brown shaded histogram) and
 \rosat-undetected (black continuous line).
}
 \label{fig:radiodist}%
\end{figure*}

The importance of the radio emission in a source is generally described using the  radio-loudness parameter. 
In this paper, we adopt the definition with the reference band in the UV, a range less affected by the host galaxy
contribution
\citep{stocke92}: 
   \begin{equation}\label{eq:rloud}
      \mathcal{R}\equiv\frac{\pedix{F}{5\,GHz,\,rf}}{\pedix{F}{2500\,\AA,\,rf}}
   \end{equation}
instead of the classical ratio with respect to optical 
$B$-band flux.
As done previously, the UV continuum flux was calculated using the QSO template, normalized to the continuum flux under
the \mgii\ line, as reported in the \sdss\ catalogue.
About $29$ QSOs are in common with \citet{strateva05}, where the authors calculated
luminosities by fitting \sdss\ spectra (after dereddening
and correcting for fibre inefficiencies, and also subtracting host-galaxy emission).
Comparing the UV luminosities for these sources, we find close agreement (mean difference of
$\Delta \log \pedix{L}{2500\,\mbox{\scriptsize \AA}} < 3$\%,
with a standard deviation of $0.1$).
This result reassures us on the accuracy of our approach.

For the $74$ detected sources in the \msam, as well as for the \csam,
the radio flux was obtained from the FIRST integrated flux density at $1.5\,$GHz (observed frame; {\tt
FINT}), assuming a power-law spectrum \pedix{F}{$\nu$}$\propto$\apix{\nu}{$-\alpha$} with index $\alpha=0.5$.
For the $737$ nondetected sources, from the flux limit of the FIRST survey ($\pedap{F}{1.5\,GHz,\,obs}{lim}=1\,$mJy) 
and the optical flux, we calculated an upper limit to the radio-loudness parameter.
The distributions of both $\mathcal{R}$ and \apix{\mathcal{R}}{up.lim} for the \msam\ are shown in 
Figure~\ref{fig:radiodist} ({\it left panel}).
The radio properties of the subsample falling in the FIRST fields are reported in Table~\ref{tab:radio}; the second part
of the Table contains the same quantities for the $11$ X-ray-undetected sources in the \csam.

From Figure~\ref{fig:radiodist} ({\it left panel}), it is evident that the detected sample is distributed rather 
uniformly in terms of the $\mathcal{R}$ parameter (green shaded histogram).
As thresholds, we assumed $\mathcal{R}=10$ and $\mathcal{R}=1$ \citep[values typically used in literature,
\eg][]{miller11}.
Out of the $74$ detected sources, $59$ are RL (all but two QSOs) and $15$ have a radioloudness parameter
intermediate between the two boundary values (RI, all QSOs); we have no detected RQ.
The $7$ nondetected sources with \apix{\mathcal{R}}{up.lim}$<1$ can be safely classified as RQ, while for the
$443$ radio-undetected AGN with $1<$\apix{\mathcal{R}}{up.lim}$\leq 10$ we can only exclude a RL classification.
Finally, for the $287$ undetected sources with $\apix{\mathcal{R}}{up.lim}>10$, we are not able to give a radio-loudness classification.

For the QSOs with sensitive radio measurements, P.~Francis produced sub-composites of RL and RQ\footnote{See
http://msowww.anu.edu.au/pfrancis/composite/lbqs\_rl\_comp.d and http://msowww.anu.edu.au/pfrancis/composite/nonbal\_rq\_comp.d, respectively}.
We checked the UV fluxes and our radio-loudness classification, both derived assuming the total composite QSO spectrum, using the 
two sub-composites; the results are in agreement within the errors.

The lack of
a clear dichotomy in the distribution of $\mathcal{R}$
is in agreement with the trend of recent works showing that the classical gap between RL and RQ fills up when
deeper and complete samples are considered \citep[\eg,][]{miller11}.
A different result might be obtained if the narrow distribution of upper limits to $\mathcal{R}$
(red dotted line in Fig.~\ref{fig:radiodist}, {\it left panel})
corresponds to a broader real distribution, peaking at lower values.
In particular, the radio-detected subsample is characterized by a rather high $\mathcal{R}$, median value $\sim 1200$.
This could be partially due to a selection bias: if high X-ray emission implies high enough radio emission, the X-ray 
selection can cause a loss of RQ sources.

The distribution of $\mathcal{R}$ for radio sources for which upper limits to the X-ray emission are
available would give important information to confirm or deny this hypothesis.
Unfortunately, the low number of AGN with X-ray upper limits in the \csam\ ($11/4508$) prevents us from
drawing firm conclusions, although their distribution seems consistent with that found for the \msam.
This suggestion is reinforced when we consider the match with the RASS: comparing the $\mathcal{R}$ distribution of
\rosat-detected ($501$ out of $4508$) and \rosat-undetected 
sources in the \csam, we found no statistical difference in terms of radio loudness (KS probability of
$0.11$; see Fig.~\ref{fig:radiodist}, {\it central panel}).
The same result is obtained when considering the radio-detected sources in the \msam: comparing \rosat-detected 
($13$ out of $67$) and \rosat-undetected sources, the hypothesis of
the same original population is confirmed with a KS
probability of $0.68$ (see Fig.~\ref{fig:radiodist}, {\it right panel}); this means that a detection or non-detection by 
\rosat\ does not relate to radio-loudness.
The same appears to be true (albeit with a much smaller sample) for \xmm\ hard X-ray detection or non-detection.

The wide range in redshift spanned by our sample can introduce evolutionary effects in the observed distribution of
$\mathcal{R}$ parameters.
To evaluate its importance, we divided our sample in $5$ bins of redshift, having a comparable number of sources.
We do not find clear evidence of redshift evolution in the RL population: the change in the $\chi^2$ when a constant is
replaced with a straight line (from $2.12$ to $1.35$) implies an F-test probability $\pedix{P}{F}\sim 0.28$.
On the other hand, the high number of upper limits compared with the number of detected RI does not allow us to draw firm
conclusions regarding the evolutionary pattern in this subclass. 
Again, to better investigate the evolutionary effects, we would need detections instead of upper limits.
From the present data, we conclude that the size and depth of our sample is the main reason of filling the gap between 
RL and RQ, rather than evolutionary effects.
If they exist, they must developed well after the QSO peak epoch studied here.

%
\begin{table}
\begin{minipage}[!ht]{\columnwidth}
\caption{Radio loudness properties in redshift bins.} \label{tab:zbin}             
\begin{center}          
\renewcommand{\footnoterule}{}  
{\scriptsize
\begin{tabular}{c@{\extracolsep{-0.03cm}} c@{\extracolsep{-0.01cm}} c@{\extracolsep{0.3cm}} c@{\extracolsep{0.3cm}} c@{\extracolsep{0.3cm}} c@{\extracolsep{0.3cm}} c@{\extracolsep{0.2cm}}}
 \hline\hline       
   & &  $z<0.7$ & $0.7 \leq z < 1.1$ & $1.1 \leq z < 1.5$ & $1.5 \leq z < 1.8$ & $z \geq 1.8$ \\
  (1) & (2) & (3) & (4) & (5) & (6) & (7) \\ 
 \hline                  
  ``main'' & RL & $9$ & $13$ & $18$ & $8$ & $11$ \\
   & det. RI  & $8$ & $2$ & $4$ & $0$ & $1$ \\
   & $\mathcal{R}^{up.lim}\in (1,10]$ & $48$ & $96$ & $114$ & $94$ & $91$ \\
   & RQ  & $3$ & $0$ & $2$ & $1$ & $1$ \\
 \hline
  ``control'' & RL & $827$ & $772$ & $953$ & $708$ & $536$ \\
   & det. RI  & $200$ & $171$ & $138$ & $98$ & $95$ \\
   & RQ  & $3$ & $4$ & $0$ & $1$ & $2$ \\
 \hline
\end{tabular}
}
\end{center}          
{\footnotesize   
$(1)$~Sample. 
$(2)$~Radio loudness classification.
$(3)$-$(7)$~Redshift bins.
} \\
\end{minipage}
\end{table}

\subsection{Eddington ratios vs radio-loudness}\label{sect:eddrrl}

\begin{figure}
 \centering
 \resizebox{\hsize}{!}{\includegraphics{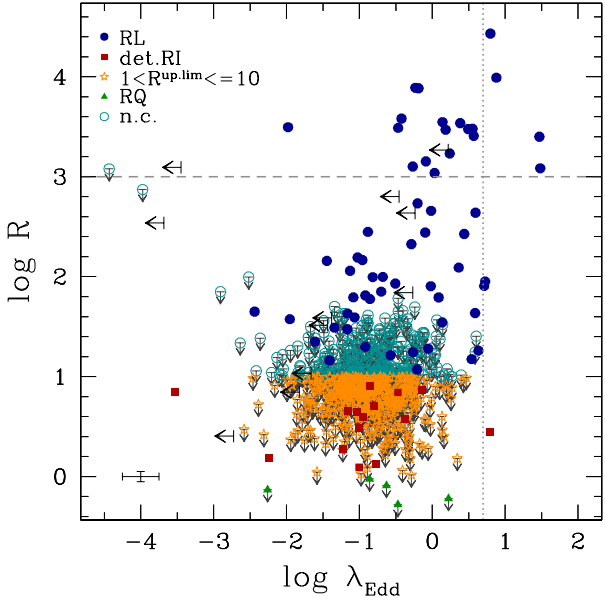}}
 \caption{Radio-loudness parameters vs. Eddington ratios.
   Key as in Fig.~\ref{fig:chklbol}:
   RL, blue filled circles; detected RI, red filled squares; undetected sources with $\mathcal{R}^{up.lim}\in (1,10]$, 
   yellow open stars; RQ, green filled triangles; not classified, light-blue open circles.
   Undetected sources are also marked with small grey arrows.
   Only mean error bars are reported to avoid clutter.
   Big black arrows correspond to the upper limits found for the $11$ X-ray non-detected sources in the \csam.
   The vertical dotted line marks the threshold of $\pedix{\lambda}{Edd} = 5$, while the horizontal dashed line 
   corresponds to $\mathcal{R}=1000$.}
 \label{fig:reddr}%
\end{figure}

Comparing Eddington ratios and radio-loudness for the sources falling in the FIRST survey area, we see an apparent 
trend of increasing $\mathcal{R}$ with $\pedix{\lambda}{Edd}$ (see Fig.~\ref{fig:reddr}).
We checked the significance of the correlation using both a generalized Kendall rank correlation test ($\pedix{\tau}{k}>0.1$) 
and a Spearman rank correlation test ($\pedix{\rho}{s}>0.15$), finding a probability $\mathcal{P}$ lower than $0.1$\% in both cases that 
a correlation is present only due to chance.
We used the {\sc ASURV} package \citep[Astronomy Survival Analysis]{feigelson85,isobe86}, that facilitates a correct 
statistical analysis when censored data (upper limits in this case) are present, although the errors are not considered 
in the calculations.
This positive correlation is in contrast with previous results, that found an increase
of radio loudness with decreasing Eddington ratio \citep[\eg,][]{ho02,merloni03,nagar05,sikora07}.
We note, however, the different range covered, narrower both in $\mathcal{R}$ and $\pedix{\lambda}{Edd}$ in our sample 
than in previous works \citep[\eg,][$\log \mathcal{R}$ between $-2$ and $7$, and $\log\pedix{\lambda}{Edd}$ between
$-6.6$ and $0.7$; cf. their fig.~3]{sikora07}.

The higher Eddington ratios observed for sources more RL could either denote a difference in the accretion mechanisms 
powering AGN with different radio-loudness, or can be related to our assumptions when calculating \pedix{L}{bol}.
Note that using a constant X-ray bolometric correction, $\pedix{\kappa}{2-10\kev}=50$, instead of a luminosity-dependent
one, the distribution of $\pedix{\lambda}{Edd}$ would become a little narrower but without changing in any significant way our results.

%

In principle, in RL sources the observed X-ray flux could be ``contaminated'' by contribution from a kpc-, or even 
pc-scale X-ray jet emission, not resolvable with \xmm, making problematic to recover the
accreting bolometric luminosities.
The fraction of X-ray emission in RL AGN that is from the jet, is a strongly debated issue.
The spurious non-nuclear emission is typically observed to be only a few percent as bright as the X-ray core
\citep[\eg,][]{marshall05}, which would not significantly change
the calculated \pedix{L}{2-10\kev} values.
Figure~\ref{fig:chklbol} confirms these considerations, suggesting that the 
overestimate of \pedix{L}{bol} due to jet contamination is relatively low. 

Otherwise, the spectral energy distribution (SED) may be different in RL and RQ
AGN, with a coronal
contribution much more important with respect to the optical emission in the former than in the latter.
In this case, applying a \pedix{\kappa}{2-10\kev} bolometric correction obtained mainly from RQ AGN would overestimate 
the optical emission, leading to a bolometric luminosity too high.
Indeed, because of these difficulties in determining the true SED linked to the accretion disk, most studies of
the X-ray bolometric correction have been performed excluding RL sources from the study of the nuclear properties.
In the sample studied by \citet{vasudevan07}, RL objects occupy the region of higher hard X-ray luminosities and lower
X-ray bolometric correction (see their fig.~3).
For one object, 3C~273, the authors are able to estimate the importance of the jet contribution, comparing old \asca\
data with new \xmm\ observations taken at a historic jet minimum.
Interestingly, despite an increasing of \pedix{\kappa}{2-10\kev} that moves the object in the upper-left direction in 
the \pedix{\kappa}{2-10\kev} vs. \pedix{L}{2-10\kev} diagram, 3C~273 still falls significantly below the 
\citet{marconi04} relation.

In most cases, the quality of the X-ray data does not allow us to perform a detailed spectral analysis.
We limited ourselves to check whether the hardness ratios ($HRs$) of the two classes are compatible 
with a steeper, accretion-dominated spectrum or with a flatter spectrum; the latter could be due to a jet-dominated
emission, or to a different disk-corona structure.
This analysis allows us to investigate, at least from a statistical point of view, also the presence of intrinsic
absorption, trying to quantify its effects. 

We considered $HRs$ defined from both soft and hard bands, dividing the \msam\ into RL 
($59$) and non-RL
($465$; see Table~\ref{tab:sum});
the $287$ non-classified sources are excluded from the $HR$ analysis.
The hard hardness ratio, $\pedix{HR}{hard}\equiv C[2-4.5]/C[4.5-12]$ (where $C[2-4.5]$ and $C[4.5-12]$ are the 
vignetting-corrected count-rates between $2$ and $4.5\,$keV observed frame, and between $4.5$ and $12\,$keV 
observed frame, respectively), less 
affected by absorption, is more directly linked with nuclear emission than the $HR$ constructed from the whole X-ray 
band.
On the other hand, comparing $\pedix{HR}{tot}\equiv C[0.2-2]/C[2-12]$ (where $C[0.2-2]$ and
$C[2-12]$ are the vignetting-corrected count-rates between $0.2$
and $2\,$keV observed frame, and between $2$ and $12\,$keV observed frame, respectively) can provide information on  the
possible presence of absorption.
Indeed, the spectral modification induced in a typical AGN spectrum by matter absorbing the 
nuclear emission, can mimic, in the total hardness ratio, a flatter (or even inverted)
power law.

%
\begin{figure}
 \centering
 \resizebox{\hsize}{!}{\includegraphics{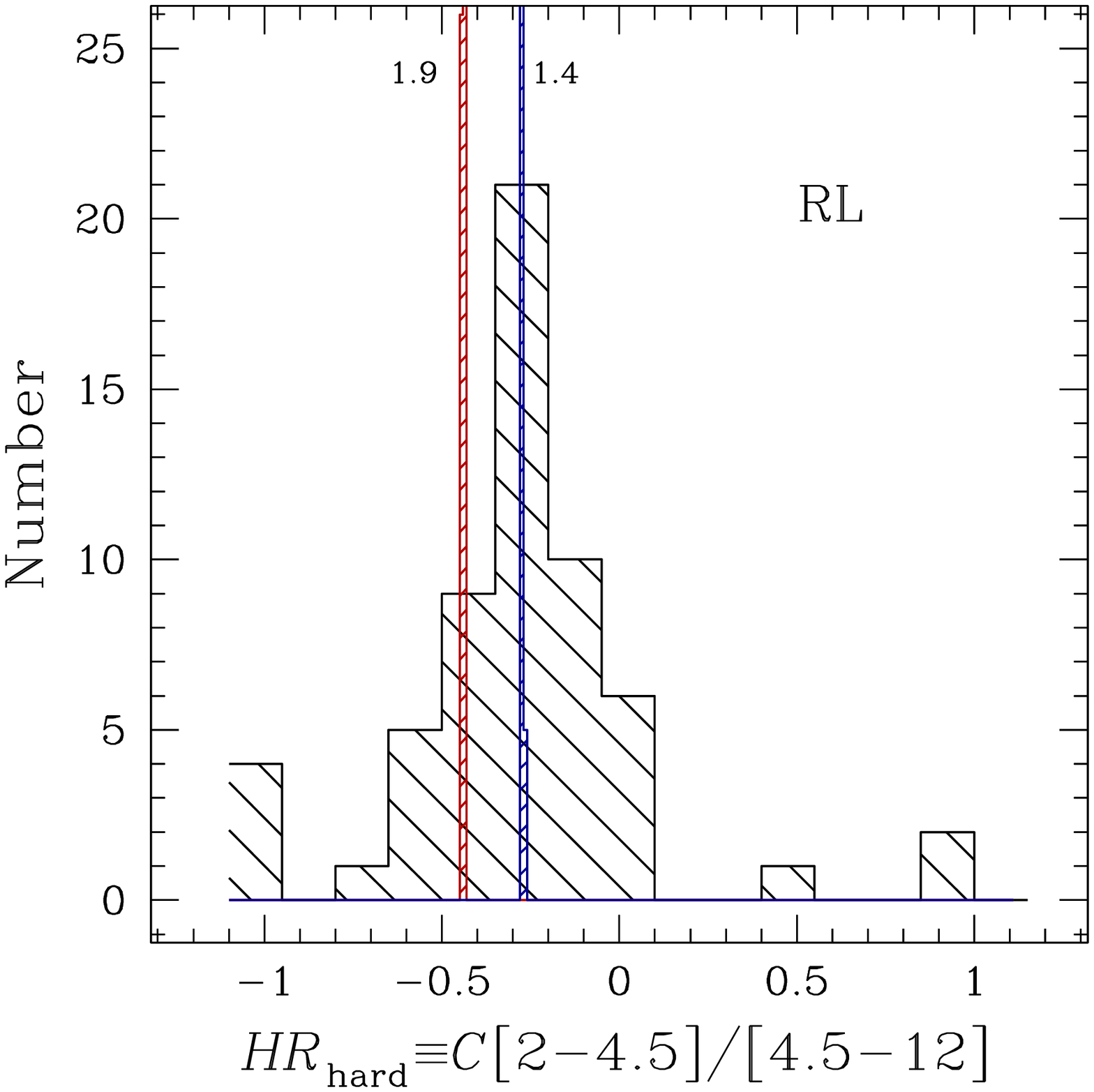}
 \includegraphics{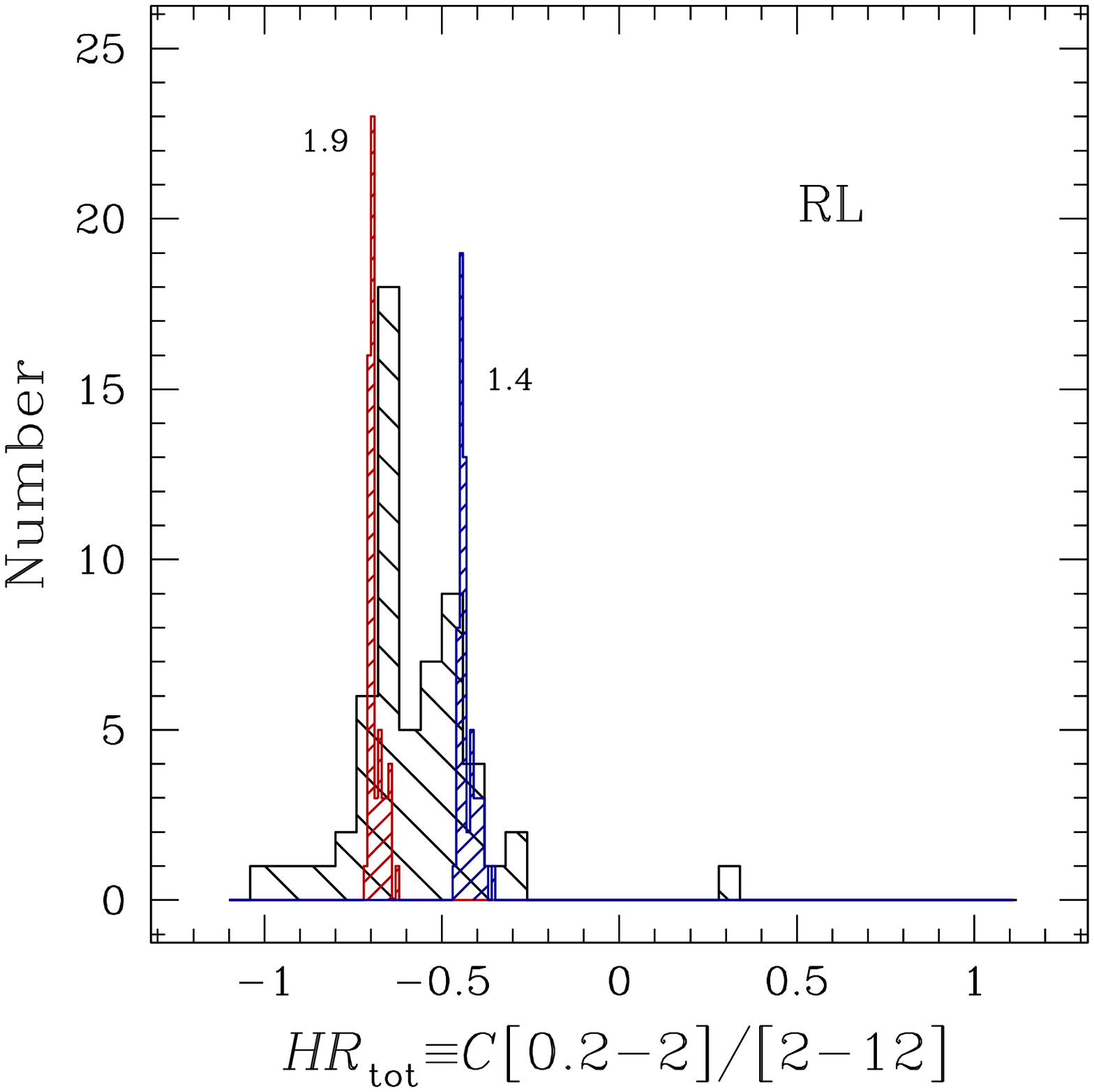}}
 \resizebox{\hsize}{!}{\includegraphics{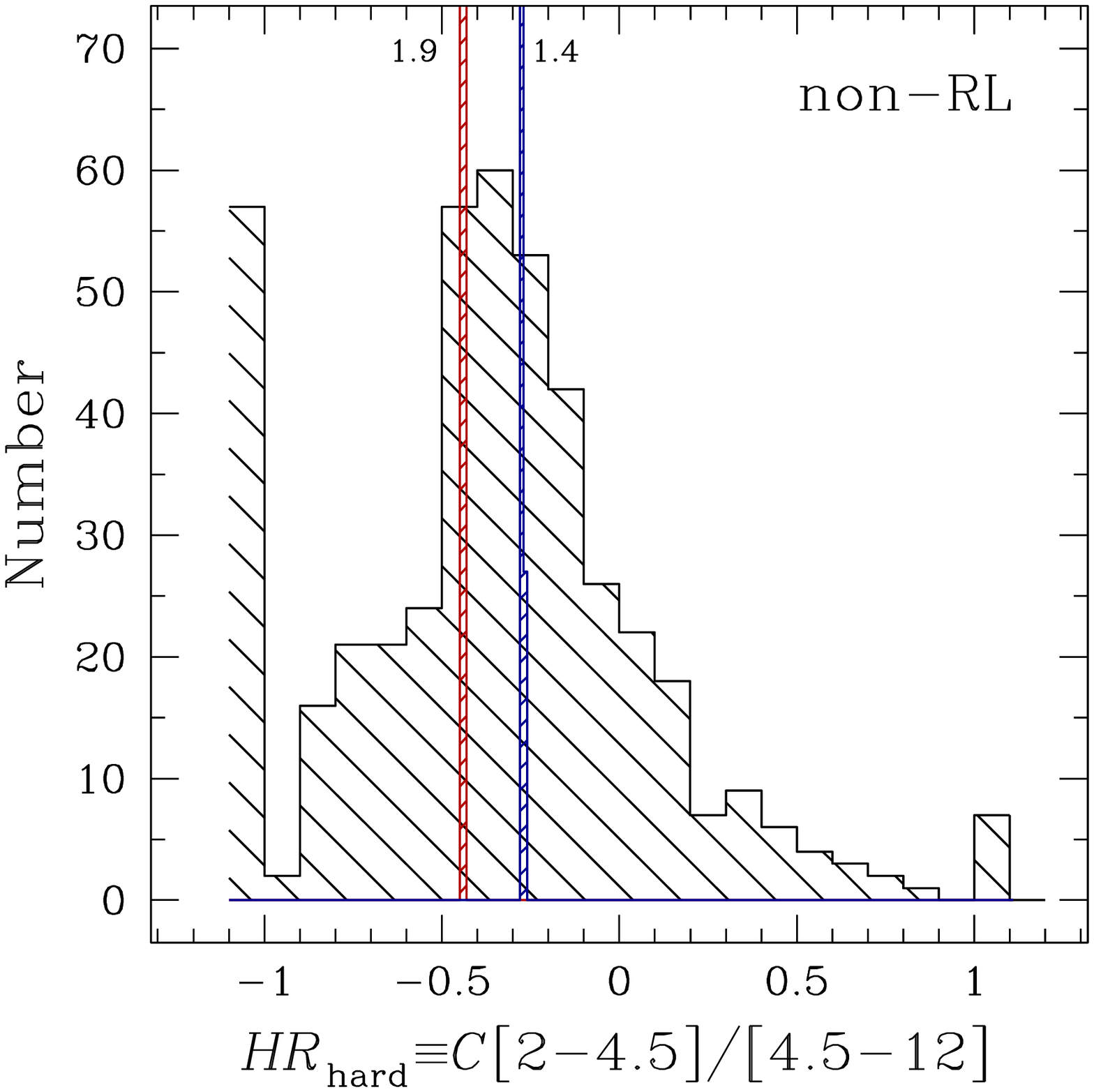}
 \includegraphics{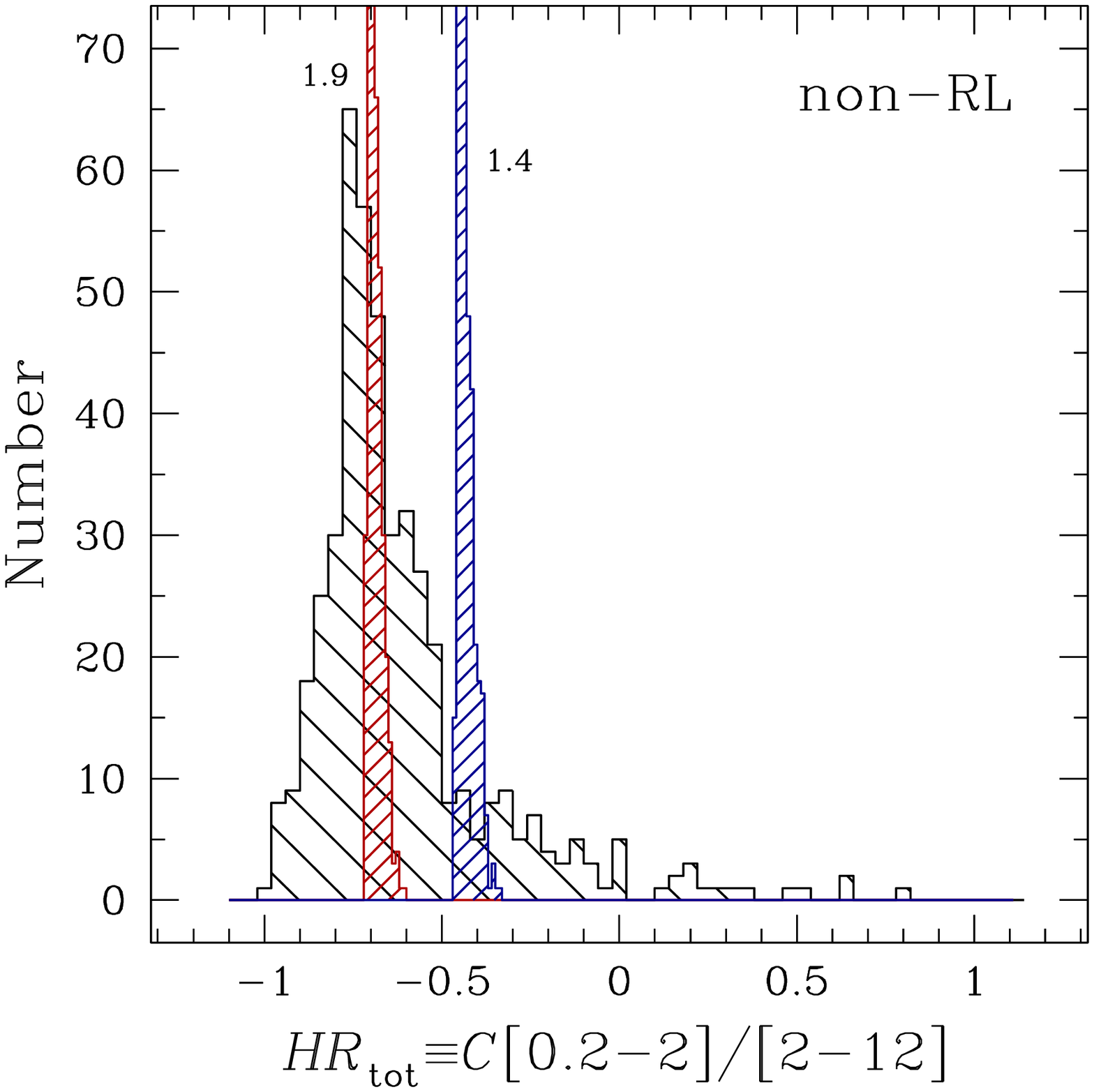}}
 \caption{Distributions of hard ({\it left panels}) and total ({\it right panels}) hardness ratios.
 Observed {\it HRs}: black shaded histograms. 
 As a comparison, we show the {\it HRs} expected for jet-dominated ($\Gamma = 1.4$, blue shaded histogram), and
 disk-dominated emission ($\Gamma = 1.9$, red shaded histogram).
 The {\it HRs} distributions have been calculated for RL ({\it upper panels}) and non-RL sources 
 (excluding the non-classified objects; {\it lower panels}).}
 \label{fig:hr}%
\end{figure}

In Figure~\ref{fig:hr} we show the observed distributions for RL ({\it upper panels}) and non-RL 
({\it bottom panels}) 
sources (hard $\pedix{HR}{hard}$, {\it left panels}; total $\pedix{HR}{tot}$, {\it right panels}).
From a KS test, the probability of the RL and non-RL subsamples being drawn from the same population is lower 
than $0.5$\% in terms of both $HRs$.
We then compared the observed $HRs$ with that expected from a jet-dominated or a disk-dominated 
emission (power law with $\Gamma = 1.4$ or $\Gamma = 1.9$, respectively), to investigate whether the X-ray colours 
reflect some difference in the components contributing to the observed emission.
The distributions expected for the \msam\ assuming a simple power-law emission covered by Galactic
absorption are overplotted to the observed ones in Figure~\ref{fig:hr}.
Note the slight broadening of the expected distributions of $\pedix{HR}{tot}$, due to the
absorption of our Galaxy and to the different redshifts of the sources.

The dispersion in the distributions is too large to allow us to draw firm conclusions.
We note, however, that the $\pedix{HR}{hard}$ suggest flatter photon indices for RL sources than
for non-RL sources: $\langle\pedap{HR}{hard}{RL}\rangle=-0.27 \pm 0.37$ and 
$\langle\pedap{HR}{hard}{non-RL}\rangle=-0.34 \pm 0.42$, to be compared with $\pedix{HR}{hard,\,1.9}\sim-0.44$
and $\pedix{HR}{hard,\,1.4}\sim-0.27$.
On the other hand, $\pedix{HR}{tot}$ is not consistent with a strong jet contribution for both RL
and non-RL sources: $\langle\pedap{HR}{tot}{RL}\rangle=-0.58 \pm 0.18$ and 
$\langle\pedap{HR}{tot}{non-RL}\rangle=-0.62 \pm 0.27$, while $\pedix{HR}{tot,\,1.4}\sim-0.43$.

For $\sim 50$\% of the non-RL subsample ($238$ sources), the analysis of the $\pedix{HR}{tot}$ suggests
a photon index flatter than $1.9$;
we investigated the possible presence of absorption in this subsample.
For each source, we calculated the expected $\pedix{HR}{tot}$ for 
an intrinsic power-law emission with $\Gamma = 1.9$, covered by a distribution of
matter with \nhsym\ spanning from $10^{21}\,$\nh\ to $10^{24}\,$\nh.
The observed ratios can be reconciled with the assumption of $\Gamma = 1.9$ assuming a column density
lower than $10^{22}\,$\nh\ for $\sim 63$\% of the $238$ sources.
Interestingly, when the same exercise is performed for the whole subsample of RL + non-RL objects, all but two\footnote{An AGN not covered by FIRST 
and a source for which the FIRST UL does not allow to obtain a radio classification.} out of the $13$ sources for which we found 
$\nhsym \geq 10^{23}\,$\nh\ are non-RL;
as
anticipated, however, the underestimate of the X-ray luminosity implied by our general assumption of unabsorbed
power-law emission is lower than a factor
of $3$, not enough to affect the apparent trend observed in Figure~\ref{fig:reddr}.

Finally, we explored the nature of the most extreme sources, with higher $\mathcal{R}$ and/or $\pedix{\lambda}{Edd}$.
Extremely beamed objects, in particular objects with featureless optical spectra,
are mostly excluded from our sample by the optical selection
criteria.
However, in this range of radioloudness and/or accretion efficiency the contamination of Flat Spectrum Radio Quasars 
(FSRQs; blazars with the optical spectrum not totally
swamped by the jet and showing broad lines produced in the BLR) can be important, implying a strong contribution of the
jet mainly in the X-ray band.
Moreover, the overestimate of the X-ray emission would be amplified (due
to the luminosity-dependent X-ray bolometric correction adopted) in calculating \pedix{L}{bol}.
From the literature, among the $7$ sources with $\pedix{\lambda}{Edd}>5$, we found $3$ blazars, $1$ FR~II, and $1$ 
lensed 
QSO (lensing effects can make difficult to correctly estimate intrinsic luminosities); no
information was found for the remaining $2$ objects.
The $11$ out of $19$
(all but $1$ QSOs) radio-detected sources
with $\mathcal{R}> 1000$, for which we found some information in the literature, are indeed
classified as blazars or FR~IIs.
In the following discussion, we will exclude these $22$ extreme sources, \ie\ those with extreme $\mathcal{R}> 1000$
and/or $\pedix{\lambda}{Edd}>5$.

The exclusion of the extreme sources reduces the probability of a correlation between $\mathcal{R}$ and $\pedix{\lambda}{Edd}$ (from a 
generalized Kendall rank correlation test: $\pedix{\tau}{k}>0.04$, $\pedix{\mathcal{P}}{k}=1.6$\%; from a Spearman rank correlation test: 
$\pedix{\rho}{s}>0.03$, , $\pedix{\mathcal{P}}{s}=29$\%), suggesting an important effect of the most strongly jetted sources in drawing 
the apparent trend observed in Figure~\ref{fig:reddr}.
 
\subsection{SED shape and luminosity-dependence}\label{sect:sed}

%
\begin{figure}
 \centering
 \resizebox{\hsize}{!}{\includegraphics{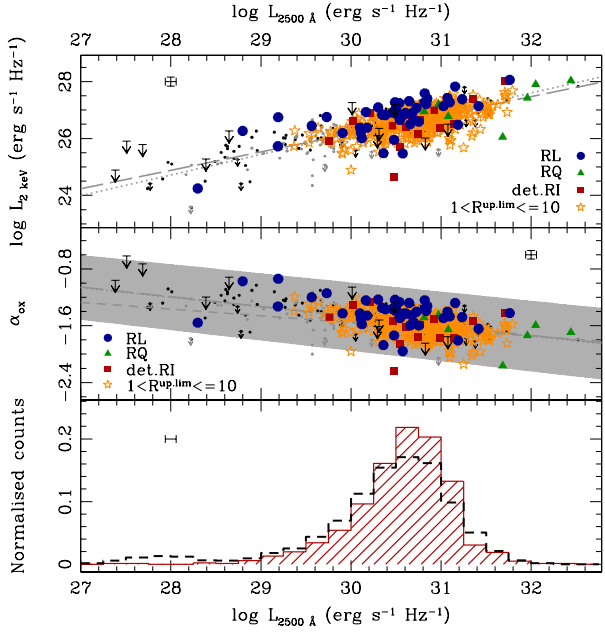}}
  \caption{Dependence on the $2500\,$\AA\ monochromatic luminosity of  \pedix{L}{2\kev} ({\it upper panel}) and  \pedix{\alpha}{ox}
  ({\it central panel}).
   Key as in Fig.~\ref{fig:chklbol}:
   RL, blue filled circles; detected RI, red filled squares; undetected sources with $\mathcal{R}^{up.lim}\in (1,10]$, 
   yellow open stars; RQ, green filled triangles.
   Big arrows correspond to the upper limits found for the $11$ X-ray non-detected sources in the \csam.
   Only mean error bars are reported to avoid clutter.
   Our data are compared with results found in the literature for samples of sources of different luminosities from the
   \sdss.
   Long-dashed lines are the best--fit linear relations for the samples by \citet{strateva05}, updating the work of
   \citet[dashed lines]{vignali03}; the dotted line represents the best fit to the \citet{steffen06} sample.
   The grey-shaded area gives the spread in the \citet{steffen06} best fit.
   Black and grey filled circles mark the \sdss\ objects with $0.1\lesssim z \lesssim 4.5$ from 
   \citet{strateva05} and the data from 
   \citet[extending their work to a larger range in luminosities]{steffen06}, respectively; arrows indicate
   upper limits in the X-ray detection.
   {\it Lower panel}:  normalized distribution in \pedix{L}{2500\ang} (red shaded histogram); the black dashed line shows the
  distribution for the \csam. }
  \label{fig:luvVSstra}%
\end{figure}

In the last years, several works have been published exploring the relation
between the emission at different energies in AGN,
as a possible indication of the mechanisms in action in samples spanning different ranges in their observational
properties.
In particular, many authors have investigated the relation between restframe UV and soft X-ray AGN emission, and its
dependence with redshift and/or optical luminosity; note that RL sources
are removed from the samples studied in most of these works.
Most studies have concluded that there is no evidence for a redshift dependence, while the X-ray emission (\ie, the
fraction of power in the accretion disk corona)
is correlated with the UV emission, and the ratio of the monochromatic X-ray to UV luminosities,
$\pedix{\alpha}{ox}\equiv \log (\pedix{F}{2\kev}/\pedix{F}{2500\ang})/\log (\pedix{\nu}{2\kev}/\pedix{\nu}{2500\ang})$,
decreases as the UV emission increases \citep{vignali03,strateva05,steffen06,just07,gibson08}.
Our sources nicely fall along the mentioned correlations, as shown in Figure~\ref{fig:luvVSstra}, with a scatter
consistent with the spread in their best fits.
Nevertheless, RL 
in general show an \pedix{\alpha}{ox} higher than expected, while detected RI lie around the regression line.
Since the luminosity dependence in the \citet[][]{marconi04}
X-ray bolometric correction is implemented via the
luminosity dependence of the \pedix{\alpha}{ox} spectral index, this result can affect our estimate of the \pedix{L}{bol} values.
However, the good agreement among the X-ray-based and the optical/UV-based estimates of \pedix{L}{bol} (see Figure~\ref{fig:chklbol} and the discussion at the end 
of Sect.~\ref{sect:lbol}) suggests that the effect in our calculation is not significant.

The different weight of the X-ray emission with respect to the UV one, again, suggest a different spectral shape for 
sources with important radio emission.
Although a contribution to the high-energy luminosity due to the jet (at least, higher than to the optical one), cannot
be completely ruled out, the exclusion of the most extreme, jetted sources from the comparison shown in
Figure~\ref{fig:luvVSstra} weakens this hypothesis.
Note that the UV continuum is often found to be redder in RL than in RQ \citep[\eg,][]{ivezic02,labita08}.
In such a scenario, a larger \pedix{\alpha}{ox} in RL objects can be obtained by assuming a reduced UV continuum emission.

Looking at  Figure~\ref{fig:luvVSstra}, it is quite evident that $5$ out of the $11$ X-ray non-detected sources in the \csam\ 
(large upper limits in the upper and central panels) have 
\pedix{L}{2500\ang} significantly lower than the luminosity observed for the sources in the \msam.
Since a significant difference between both samples in the UV luminosity would affect our analysis,  we checked that these low luminosities are not representative 
of the values for the whole \csam.
A comparison between the distributions in \pedix{L}{2500\ang} of the two samples (see Figure~\ref{fig:luvVSstra}, {\it lower panel}) demonstrates 
that the two samples cover similar ranges in UV luminosity, both 
peaking at $\sim 3\times10^{31}\,$\lumHz, with the distribution of the \csam\ extending down to lower luminosities, where about half of the X-ray non-detected 
sources can be found.

Comparing the radio-loudness with respect to different energy ranges (\eg,
optical, UV, X-ray) can provide hints on the relations between the emission in
the different bands, and/or their evolution with redshift or luminosity.
In the following, we consider the radio loudness
parameter with respect to the hard X-ray
luminosity \citep{terashima03}:
   \begin{equation}\label{eq:xloud}
      \pedix{\mathcal{R}}{X}\equiv\frac{\pedix{{\nu}L}{5\,GHz,\,rf}}{\pedix{L}{2-10\kev,\,rf}}
   \end{equation}
The values of \pedix{\mathcal{R}}{X} (or their 
upper limits, for radio-undetected sources) are reported in
Table~\ref{tab:radio}; 
in the second part, we report the lower limits to radio-to-X-ray ratio for the 
AGN in the \csam\ for which we have X-ray upper limits.

%
\begin{figure}
 \centering
 \resizebox{\hsize}{!}{\includegraphics{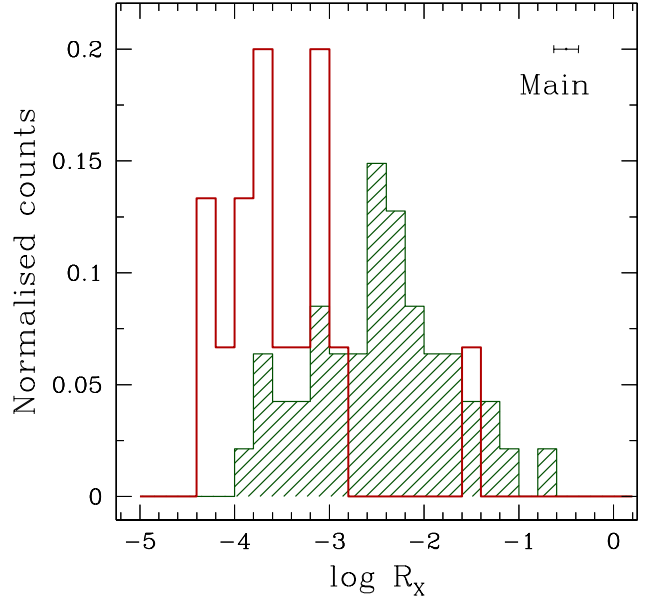}}
 \caption{Normalised distribution of \pedix{\mathcal{R}}{X} for the radio-detected sources in the \msam.
 RL, green shaded histogram; detected RI, red continuous line.
 The error bar reported represents the mean uncertainties in \pedix{\mathcal{R}}{X}, obtained as detailed in
 Sect.~\ref{sect:err}.
}
 \label{fig:xrdist}%
\end{figure}

The distribution of $\pedix{\mathcal{R}}{X}$ for detected sources suggests the possible presence of a gap at
$\log\pedix{\mathcal{R}}{X}\sim -3$.
However, whether it is real or not depends again on the distribution of undetected sources.
All the RL sources (\ie, sources with $\mathcal{R} > 10$) have $\log \pedix{\mathcal{R}}{X}> -4$;
the distinction between RL and RI tends to disappear, having $12$ out of $15$ detected RI $\pedix{\mathcal{R}}{X}$
higher than this threshold (see Fig.~\ref{fig:xrdist}).

The dependence of these boundaries with the luminosity is still an open issue.
For a sample of local Seyfert galaxies and low-luminosity radio galaxies,
\citet{panessa07} found for RL/non-RL separating values of $\log
\pedix{\mathcal{R}}{X}\sim -2.8$ and $\log \pedix{\mathcal{R}}{4400}\sim 2.4$, 
while the addition of luminous PG quasars implies a $\log 
\pedix{\mathcal{R}}{X}\sim -4.5$, fixing  $\pedix{\mathcal{R}}{4400}=10$ to 
define RL sources \citep{terashima03}.

Note that the radio-to-optical loudness was calculated by \citet{terashima03}
and \citet{panessa07} using the optical luminosity at $4400\,$\AA\ instead of
the UV luminosity at $2500\,$\AA.
Therefore, to properly compare the our \msam\ and the low-luminosity sample by
\citet{panessa07}, we have to evaluate the same
parameter, \pedix{\mathcal{R}}{4400}.
Again, to estimate the optical luminosity at such wavelength, we adopted the QSO
template, normalized to the continuum flux under the nearest feature to this wavelength, 
in this case the \hb\ line, as reported in the \sdss\ catalogue.
This means that a comparison can be performed only for the $185$ sources showing
this line in their \sdss\ spectra.
As before, for the radio nondetected sources we calculated an upper limit to the
optical radio-loudness parameter from the
flux limit of the FIRST survey.
We note that our approach could possibly overestimate the $4400\,$\AA\  flux by including host galaxy emission, 
that can contributes to the optical band, in a particularly significant way  for AGN with lower optical luminosity.
In these sources, the resulting \pedix{\mathcal{R}}{4400} values come out smaller than the real one.

\begin{figure*}
 \centering
 \resizebox{\hsize}{!}{
 \includegraphics{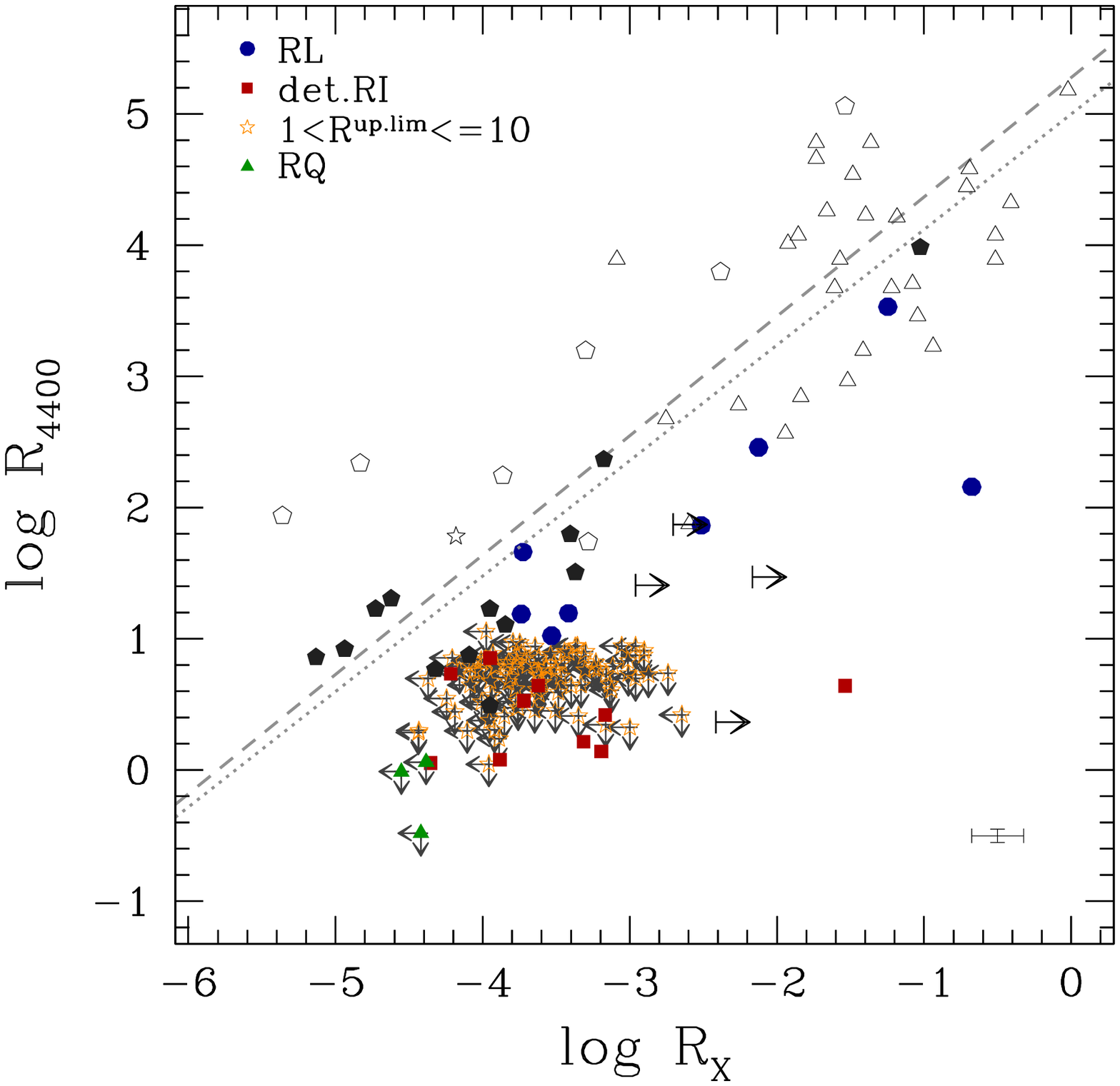}
 \includegraphics{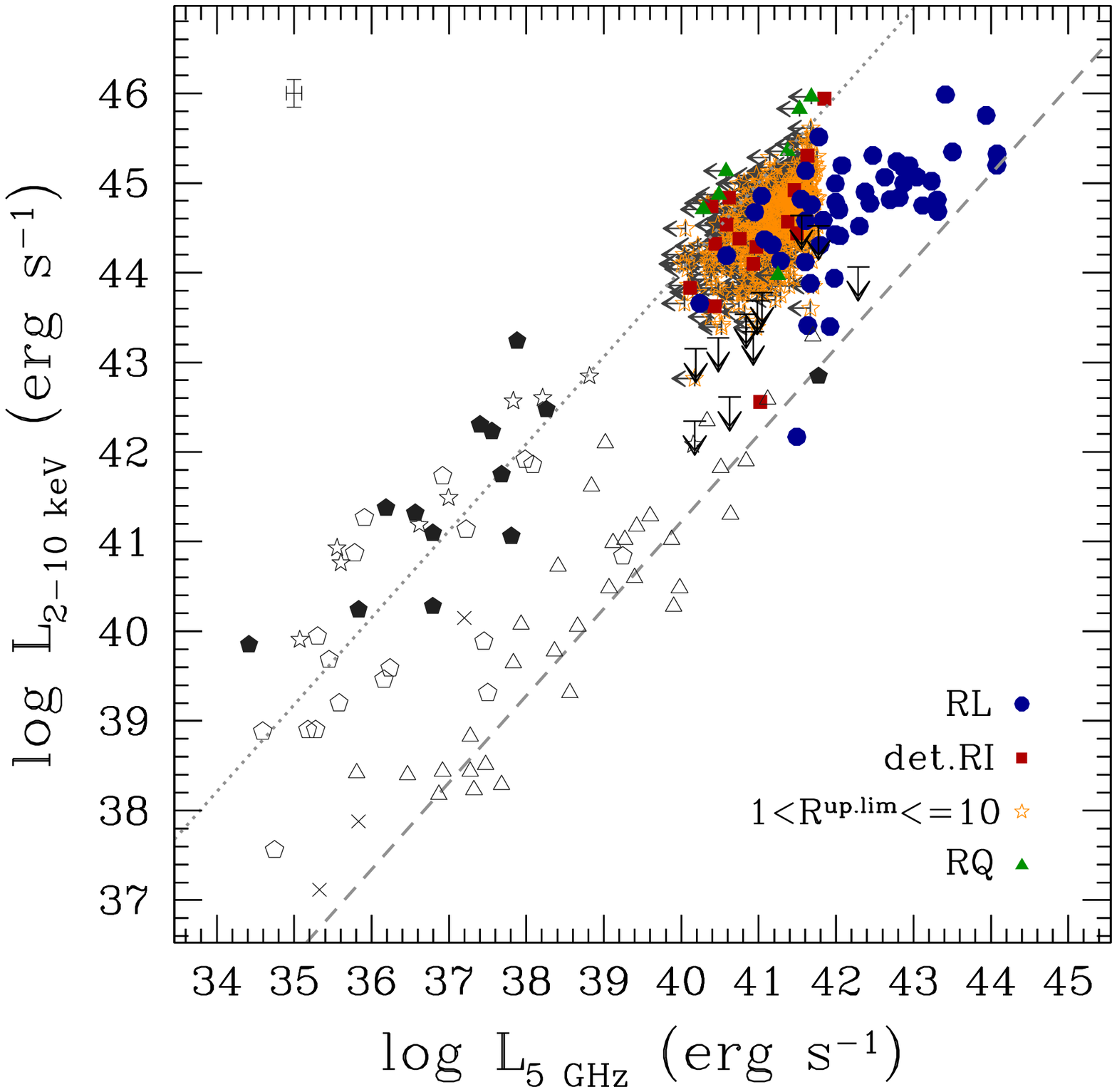}
  \includegraphics{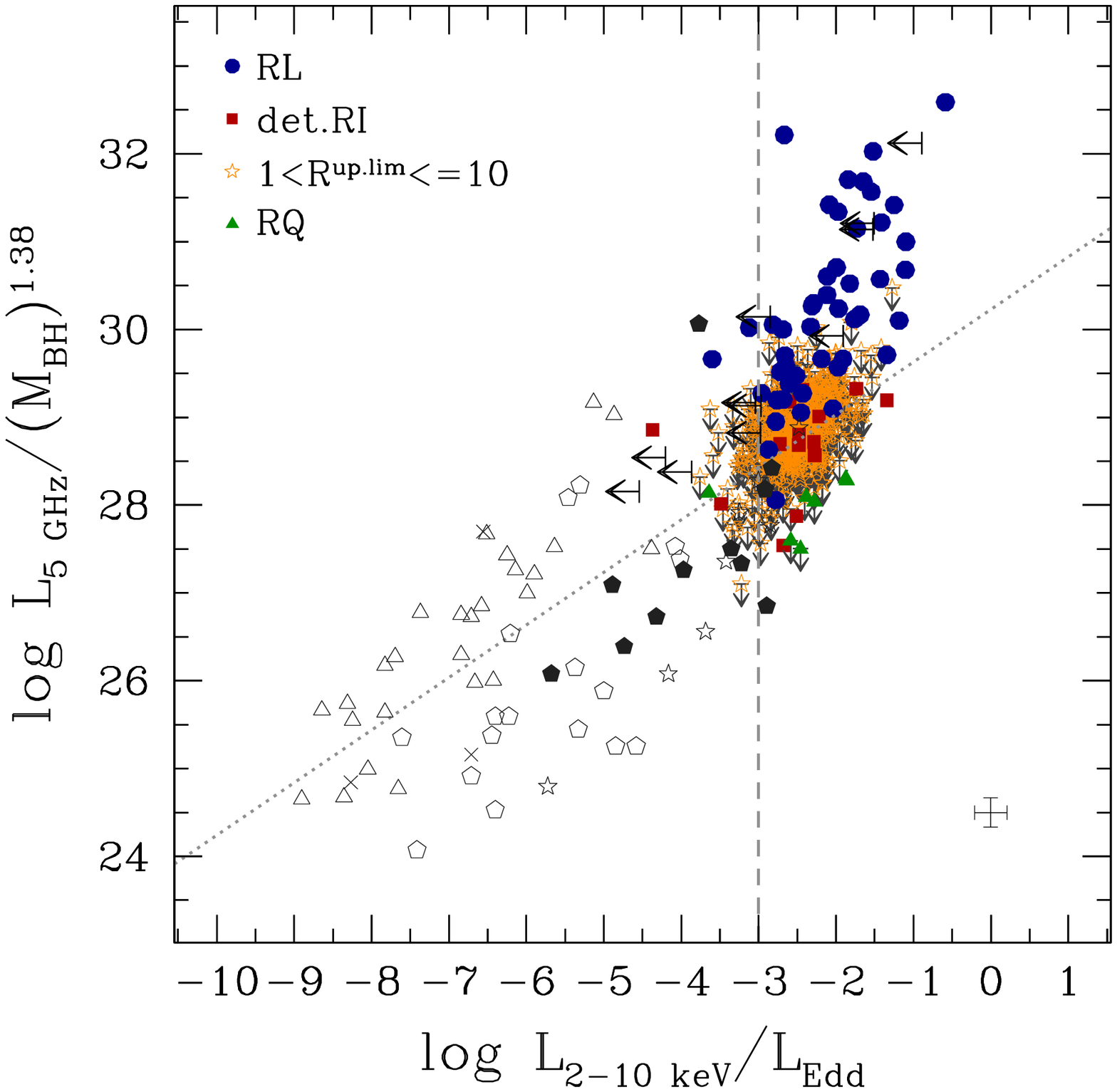}}
  \caption{
   {\it Left panel:} Relation between the radioloudness parameters \pedix{\mathcal{R}}{4400} and 
  \pedix{\mathcal{R}}{X}; dotted and dashed grey
   lines are the best-fit regression lines found by \citet{terashima03} and \citet{panessa07}, respectively.
   {\it Central panel:} radio vs. X-ray luminosity; dotted and dashed grey
   lines are the best-fit regression lines found by \citet{panessa07} for
   Seyfert and low-luminosity radio galaxies,
   respectively. 
   {\it Right panel:} X-ray--to--Eddington luminosity ratio compared to the AGN fundamental plan; dotted line 
   corresponds to the fundamental plane equation, by \citet{merloni03}, while vertical dashed line marks the
   $\pedix{L}{X}/\pedix{L}{Edd}$ where the switch in the accretion mode is expected.
   Key as in Fig.~\ref{fig:chklbol}:
   RL, blue filled circles; detected RI, red filled squares; undetected sources with $\mathcal{R}^{up.lim}\in (1,10]$, 
   yellow open stars; RQ, green filled triangles.
   Undetected sources are also marked with small grey arrows.
   Only mean error bars are reported to avoid clutter.
   Big black arrows correspond to the upper limits found for the $11$ X-ray non-detected sources in the \csam.
   Our data are compared with the results found by \citet{panessa07} for a sample of low-luminosity AGN (grey symbols):
   filled and open polygons, Seyfert~1 and 2, respectively; open stars, Compton-thick candidates; crosses, ``mixed
   Seyfert''; open triangles, low-luminosity radio galaxies.}
  \label{fig:xrFP}%
\end{figure*}

The results are reported in Table~\ref{tab:radio}; the second part
of the Table contains the same quantities for the $11$ X-ray-undetected sources
in the \csam.
The $\log \pedix{\mathcal{R}}{4400}$ versus $\log \pedix{\mathcal{R}}{X}$
is shown in Figure~\ref{fig:xrFP} ({\it left panel}); in the same plot we report
also the \citet{panessa07} sample.
Clearly, our sample falls below both correlations found by \citet{terashima03}
and \citet{panessa07}; this is observed apart from the classification in terms
of $\mathcal{R}$, although the effect is more evident for sources with lower
$\mathcal{R}$.

%

The main result we can draw is that the increase in the emission in the X-ray
and radio bands seems to proceed in a linked way: higher X-ray luminosity
corresponds to higher radio luminosity, so that the range spanned in 
\pedix{\mathcal{R}}{X} is roughly the same.
On the contrary, the optical luminosity changes independently, producing the
observed decrease in \pedix{\mathcal{R}}{4400}.
We note that the large fraction of radio upper limits in our sample can partially
affect all the considerations we are doing: detections instead of upper limits
could in principle change the overall distribution of our sample in the
different distribution diagrams.

A possibility is that the former connection is just apparent, mainly due to a
strong contamination of the X-ray observed emission due to the jet, \ie\ the
same physical component where the radio emission is produced.
However, in the previous sections we excluded the presence of a significant
contribution of X-rays from the jet.
Moreover a direct comparison between the $2-10\,$keV
luminosity and the radio luminosity at $6\,$cm demonstrates that this is not the
case (see Fig.~\ref{fig:xrFP}, {\it central panel}).
As expected, we are sampling a range of
higher luminosities both at short and long wavelengths.
At lower and intermediate radio luminosities our sample falls along the best-fit
line found by \citet{panessa07} for
their Seyferts (dotted line).
More interestingly, an effect of ``saturation'' seems to take place, with the
sources at higher luminosities (mainly RL)
shifting towards the locus of low-luminosity radio galaxies.
This result reinforces our former conclusion that the X-ray
luminosity is not jet-dominated, even in RL sources,
confirming in addition our X-ray--based estimation of nuclear properties.
Therefore, we are left with the only conclusion that we are looking at a real
different relation between the jet and (on one side) the regions where X-rays are emitted and (on the other side) the regions where optical radiation is produced.
While the presence of  jet changes the structure of the BH-accretion disk-corona systems so that the X-ray emission is strongly affected, 
the effects on the observed optical emission is lower, since the contribution from the most external region of the disk basically is likely untouched.
Quite interestingly, \citet[][]{cleary07}, investigating the MIR properties in a sample of extremely powerful radio sources with \spitzer, found 
an ``upper envelope'' in the observed $15\,$\mum\ luminosity for the most powerful RL AGN.
In their analysis the authors suggest that, if observed at
higher radio lobe luminosities than those of their sample, this may provide support for the ``receding torus'' model \citep[][]{lawrence91}.
Assuming a MIR-to-X-ray ratio of $\sim10$ \citep[from the global SEDs of][]{richards06}, our observed saturation limit matches 
approximately their limit of $\pedix{L}{15\mum}=10^{24.5}\,$W~Hz$^{-1}$~sr$^{-1}$.

The accretion regime in action in our sample is clearly different from that
characterizing the \citet{panessa07} sample of Seyfert galaxies.
The ratio between X-ray and
Eddington luminosities is higher (see Fig.~\ref{fig:xrFP}, {\it right panel}), while the importance of the radio emission
with respect to the X-ray emission is comparable (see Fig.~\ref{fig:xrFP}, {\it left panel}):
although our sample spans a small range in Eddington ratios, looking at the radio-detected AGN (\ie, the blue filled circles, RL, and the 
red filled squares, detected RI, in Fig.~\ref{fig:xrFP}) it seems that we are drawing a 
parallel track in the \pedix{\mathcal{R}}{X}-\pedix{\mathcal{R}}{4400} plane, moving at higher
accretion.
Again, the true distribution of the undetected sources (upper limits in both quantities, clustered at $\pedix{\mathcal{R}}{X}\sim 10^{-4}-10^{-3}$ 
and $\pedix{\mathcal{R}}{4400}\sim1-10$) may affect our conclusions.
This important difference is confirmed when considering the so-called
``fundamental plane of black hole activity'' \citep[see Fig.~\ref{fig:xrFP}, {\it right panel}]{merloni03}: the majority of 
our AGN fall at accretion ratios higher than the threshold of
$\pedix{L}{2-10\kev}/\pedix{L}{Edd}>10^{-3}$, where the switch between
radiatively inefficient (ADAF) and radiatively efficient accretion flow is
expected to occur.
At these accretion regimes, the relation between accretion flow and
jet power changes, as demonstrated by the deviation of our sample from the
fundamental-plane equation (dotted line).

To summarize, we extend the analysis of the correlation between emission in
different bands to higher luminosities for a wider sample of
efficiently-accreting objects.
The observed trend with the luminosity of the radio, optical, and X-ray
emission and their correlations lead us to suggest that the radio emission is
strongly coupled with a non-jet-dominated X-ray radiation, produced in the 
innermost region of the SMBH-accretion disk system, instead of with the optical
one, 
which originates in a superposition of emission from material at different distances from the nucleus.


\section{Conclusions}\label{sect:concl}

In this paper, we explored the interplay between X-ray and radio emission in
type~1 AGN, in order to investigate the origin of radio emission in the
framework of the AGN Unification Model, and its relation with the different
physical components of the central system as well as with the different
accretion regimes in act.
The availability of deep catalogues at different energies, from radio up to
X-rays bands, with wide sky coverage, allow us to collect multiwavelength
information for a large sample of
$\sim 800$ type~1 AGN, spanning a redshift range from $0.3$ to $2.3$.

For all the sources, we obtained the masses of the central black hole from the
optical spectra, using the well-known relation between mass, emission-line
width, and continuum luminosity.
X-ray data were used to compute the bolometric output of the sources; being produced 
in the innermost regions of the central engine, the high-energy emission is one of the best proxies to estimate the 
bolometric luminosity, less affected by effects of reprocessing and external contamination than radiation emitted at 
larger distances.
We tested our X-ray-based estimate of nuclear properties against the possible absorption in the X-ray band, finding that
its effects would be negligible.
In the previous sections, we discussed extensively the importance of the only contaminant we expect at high energy, 
\ie\ the emission from the jet in the most powerful radio sources.

Combining SMBH masses and bolometric luminosities, we recovered the Eddington ratios; the collection of radio 
information allow us to characterize the sample in terms of importance of the radio emission in the global energetic 
output.
We note that one of the main characteristics of this work is the derivation of the different physical quantities from 
observations in different energy ranges, compared with
the unavoidable dependence expected when observations in the same energy band are used (\eg, the optical emission
adopted both in the determination of the SMBH mass and as a proxy of the bolometric luminosity).

To assess whether these conclusions are a property related to the X-ray
selection character of our sample, we tested our
conclusions against a sample of FIRST radio AGN which appear in the \sdss\
catalogue, and for which X-ray information is
obtained from the \rosat\ All-Sky Survey \citep[RASS;][]{rass}.

Below we summarize our main results:
  \begin{enumerate}
      \item Our sources have typically $\pedix{\lambda}{Edd}> 0.01$. The sample analysed, which is effectively X-ray 
      selected, might be biased towards high accretion rates.
It is not surprising then that the trend of $\mathcal{R}$ increasing towards
decreasing $\pedix{\lambda}{Edd}$ noted in
local samples is
absent in our study, as we do not expect ADAFs or other radio-prone
low-efficiency accretion modes to be present.
We also find a few extreme cases (both in terms of $\mathcal{R}$ and
$\pedix{\lambda}{Edd}$), that we identify with
beamed sources.
      \item At a variance with lower-luminosity samples, ours does not show any hint of
bimodality in radio-loudness.
Despite the fact that our sample spans a wide redshift range, we have not found compelling evidence
that bimodality develops with cosmic time.
Although this cannot be excluded, we rather believe that the absence of a
gap between RL and RQ is mainly due to the size of our sample, highly increased
with respect to previous works.
Our analysis suggests that, if the RL/RQ bimodality exist, it is a local effect.
      \item We have computed X-ray loudness \pedix{\mathcal{R}}{X}
for RL and RQ AGN (excluding strongly jetted sources) and we conclude that
in the bulk of the AGN population, radio emission is tightly linked to the
accretion disk and not to larger scale
phenomena.
   \end{enumerate}

\begin{acknowledgements}
Based on observations obtained with \xmm\ (an ESA science mission with instruments and contributions directly funded by
ESA Member States and the USA, NASA).
Funding for the \sdss\ and \sdss-II has been provided by the Alfred P. Sloan Foundation, the Participating Institutions,
the National Science Foundation, the U.S. Department of Energy, the National Aeronautics and Space Administration, the
Japanese Monbukagakusho, the Max Planck Society, and the Higher Education Funding Council for England.
The \sdss\ Web Site is http://www.sdss.org/.
The \sdss\ is managed by the Astrophysical Research Consortium for the Participating Institutions. 
This research has made use of NASA's Astrophysics Data System.
The research uses the interactive graphical viewer and editor for tabular data TOPCAT (http://www.starlink.ac.uk/topcat/) and its 
command-line counterpart STILTS (http://www.starlink.ac.uk/stilts/).
We warmly thank the referee for her/his suggestions that significantly improved the paper.
LB acknowledges support from the Spanish Ministry of Science and Innovation through a ``Juan de la Cierva'' fellowship.
Financial support for this work was provided by the 
Spanish Ministry of Economy and Competitiveness through research grant AYA2010-21490-C02-01
FJHH acknowledges support from CSIC through the undergraduate research programme ``JAE-Introducci\'on a la
investigaci\'on''.
\end{acknowledgements}

\bibliographystyle{aa} 
\bibliography{hh-ms} 


%
\begin{landscape}
\begin{deluxetable}{r c c c r c c c c c c c c c}
 \tabletypesize{\tiny}
 \tablecolumns{14}
 \tablewidth{0pt}
 \tablecaption{\footnotesize Basic information for the $852$ X-ray selected $+11$ X-ray undetected \sdss\ type~1 AGN. \label{tab:sample}}
 \tablehead{
  \multicolumn{4}{c}{2XMMi} & \colhead{} & \multicolumn{8}{c}{\sdss} & \colhead{} \\ 
  \cline{1-4} \cline{6-13} \\ 
  \colhead{SRCID} & \colhead{R.A.} & \colhead{Dec.} & \colhead{\pedix{F}{0.2-12\kev}} & \colhead{Redshift} & \colhead{specObjID} & \colhead{R.A.} & \colhead{Dec.} & \colhead{\pedix{g}{SDSS}} & \colhead{\pedix{r}{SDSS}} & \colhead{\pedix{i}{SDSS}} & \colhead{$\sigma(\mgii)$} & \colhead{\pedix{F}{c}@$2799\,$\AA\,rf} & \colhead{{\tt FINT}@$1.5\,$GHz\,obs} \\
  \colhead{(1)} & \colhead{(2)} & \colhead{(3)} & \colhead{(4)} & \colhead{(5)} & \colhead{(6)} & \colhead{(7)} & \colhead{(8)} & \colhead{(9)} & \colhead{(10)} & \colhead{(11)} & \colhead{(12)} & \colhead{(13)} & \colhead{(14)} }
\startdata
 \cutinhead{\it X-ray detected AGN}
  $192021$ & $2.5419$ & $ 0.8575$ & $20.625\pm 3.563$ & $0.3871\pm0.0011$ & $109716225189216256$ & $2.5418$ & $ 0.8574$ & $19.210$ & $18.685$ & $18.475$ & $ 9.878\pm2.875$ & $ 5.367$ &  \emph{undet.}      \\        
  $192103$ & $2.8767$ & $ 0.9644$ & $38.860\pm 2.617$ & $1.4934\pm0.0014$ & $109434742759227392$ & $2.8767$ & $ 0.9644$ & $20.411$ & $20.308$ & $20.136$ & $37.148\pm5.183$ & $ 3.405$ & $156.0718\pm0.1049$ \\
  $192150$ & $5.5413$ & $ 0.2747$ & $ 4.366\pm 1.169$ & $0.5749\pm0.0012$ & $110279119400337408$ & $5.5415$ & $ 0.2748$ & $18.043$ & $18.066$ & $17.918$ & $13.672\pm0.477$ & $21.257$ &  \emph{OutFIRST}    \\        
  $192335$ & $7.5410$ & $-0.3762$ & $ 9.835\pm13.374$ & $0.4099\pm0.0011$ & $110560641432944640$ & $7.5411$ & $-0.3760$ & $19.334$ & $19.102$ & $18.828$ & $19.375\pm1.788$ & $ 6.224$ &  \emph{undet.}     	       
  \\[0.1cm]
 \multicolumn{14}{l}{[\nodata]}
  \\[0.1cm]
  \hline \hline
  \\[0.01cm]
 \cutinhead{\it X-ray undetected AGN}
  $01$Xund & $126.6892$ & $26.5414$ & $ < 0.833$ & $0.3783\pm0.0002$ & $446646711809474560$ & $126.6892$ & $26.5414$ & $20.266$ & $19.026$ & $18.388$ & $84.222\pm49.750$ & $0.142$ & $1.3000\pm0.1310$ \\	      
  $02$Xund & $130.1848$ & $19.6289$ & $ < 1.069$ & $0.4500\pm0.0002$ & $641149183918604288$ & $130.1848$ & $19.6289$ & $21.879$ & $20.220$ & $19.486$ & $99.959\pm48.131$ & $0.908$ & $2.5000\pm0.1350$ \\	     
  $03$Xund & $131.8893$ & $34.9104$ & $ < 1.188$ & $1.0282\pm0.0020$ & $263123853125877760$ & $131.8893$ & $34.9104$ & $19.447$ & $19.148$ & $19.139$ & $23.386\pm 1.242$ & $6.918$ & $0.6400\pm0.1510$ \\    
  $04$Xund & $149.6652$ & $30.8494$ & $ <17.833$ & $0.9730\pm0.0013$ & $549105712635052032$ & $149.6652$ & $30.8494$ & $20.258$ & $19.767$ & $19.559$ & $31.507\pm 2.546$ & $4.270$ & $3.7400\pm0.1430$       
  \\[0.1cm]
 \multicolumn{14}{l}{[\nodata]}
  \\[0.1cm]
\enddata
\tablefoot{
 The complete table is published in its entirety in the electronic edition of the journal.
 A portion is shown here for guidance regarding its form and content.
 In the first part, we list the X-ray detected AGN; in the second part, we report basic data for the $11$ sources in the \csam\ undetected in the X-ray (see text).\\
 Col. (1) : Unique source identification number in the 2XMMi catalogue; for the X-ray undetected sources, arbitrary ID number.
 Col. (2) \& (3): J2000 right ascension and declination of the X-ray object (in degree).
 Col. (4): \xmm\ EPIC-pn flux between $0.2$ and $12\,$keV (entry {\tt xmm\_pn\_8\_flux} in the 2XMMi catalogue), in units of $10^{-14}\,$\flux; for the X-ray undetected sources, \xmm\ EPIC upper limit to the flux from {\tt FLIX}.
 Col. (5): Redshift for the \sdss\ counterpart.
 Col. (6): Unique spectroscopic ID in the \sdss\ catalogue.
 Col. (7) \& (8): J2000 right ascension and declination of the optical object (in degree).
 Col. (9) - (11): Magnitudes in the $g$, $r$, and $i$ \sdss\ filters.
 Col. (12): Standard deviation of the Gaussian fitted to the \mgii\ line (in \AA).
 Col. (13): Continuum value below the \mgii\ line (in units of $10^{-17}\,$\fluxA).
 Col. (14): Integrated radio flux density at $1.5\,$GHz (observed frame; in units of mJy), from the FIRST survey; \emph{OutFIRST}: source not covered by FIRST; \emph{undet.}: source covered but not detected by FIRST.}
\end{deluxetable}
\end{landscape}
%


%
\begin{landscape}
\begin{deluxetable}{r c c c c c c c c c c}
 \tabletypesize{\scriptsize}
 \tablecolumns{11}
 \tablewidth{0pt}
 \tablecaption{\footnotesize Derived properties for the $852$ X-ray selected $+11$ X-ray undetected \sdss\ type~1 AGN. \label{tab:deriv}}
 \tablehead{
  \colhead{SRCID} & \colhead{Redshift} & \colhead{\pedix{N}{H,\,gal}} & \colhead{FWHM(\mgii)} & \colhead{\pedix{\log L}{cont}@$3000\,$\AA\ rf} & \colhead{\pedix{\log M}{BH}} & \colhead{\pedix{\log L}{2-10\kev}} & \colhead{$\log\left(\pedix{L}{2-10\kev}/\pedix{L}{Edd}\right)$} & \colhead{\pedix{\log \kappa}{2-10\kev}} & \colhead{\pedix{\log L}{bol}} & \colhead{$\log\pedix{\lambda}{Edd}$}\\
  \colhead{(1)} & \colhead{(2)} & \colhead{(3)} & \colhead{(4)} & \colhead{(5)} & \colhead{(6)} & \colhead{(7)} & \colhead{(8)} & \colhead{(9)} & \colhead{(10)} & \colhead{(11)} }
\startdata
 \cutinhead{\it X-ray detected AGN}
  $192021$ & $0.3871\pm0.0011$ & $2.77$ & $1294.88\pm 519.92$ & $40.600\pm0.049$ & $7.123\pm0.350$ & $43.722\pm0.075$ & $-1.515\pm0.358$ & $1.287\pm0.029$ & $45.009\pm0.080$ & $-0.228\pm0.359$ \\
  $192103$ & $1.4934\pm0.0014$ & $2.66$ & $3392.79\pm1235.12$ & $42.009\pm0.008$ & $8.664\pm0.316$ & $45.411\pm0.029$ & $-1.367\pm0.318$ & $1.936\pm0.011$ & $47.348\pm0.031$ & $ 0.569\pm0.318$ \\
  $192150$ & $0.5749\pm0.0012$ & $2.75$ & $2163.19\pm 125.94$ & $41.610\pm0.004$ & $8.074\pm0.051$ & $43.455\pm0.116$ & $-2.732\pm0.127$ & $1.184\pm0.045$ & $44.639\pm0.125$ & $-1.549\pm0.135$ \\
  $192335$ & $0.4099\pm0.0011$ & $2.42$ & $3964.57\pm 544.14$ & $40.683\pm0.020$ & $8.136\pm0.120$ & $43.454\pm0.591$ & $-2.797\pm0.603$ & $1.183\pm0.227$ & $44.637\pm0.633$ & $-1.614\pm0.644$
 \\[0.1cm]
 \multicolumn{11}{l}{[\nodata]}
  \\[0.1cm]
  $204283$ & $0.9268\pm0.0029$ & $3.15$ & $2056.91\pm 608.64^{(\ast)}$ & $41.245\pm0.008^{(\ast)}$ & $7.848\pm0.607^{(\ast)}$ & $43.500\pm0.420$ & $-2.462\pm0.738$ & $1.201\pm0.162$ & $44.700\pm0.451$ & $-1.261\pm0.756$
  \\[0.1cm]
 \multicolumn{11}{l}{[\nodata]}
  \\[0.1cm]
  \hline \hline
 \\[0.01cm]
 \cutinhead{\it X-ray undetected AGN}
  $01$Xund & $0.3783\pm0.0002$ & $3.46$ & $15411.71\pm12547.71^{(\ast)}$ & $38.915\pm0.008^{(\ast)}$ & $8.432\pm0.896^{(\ast)}$ & $ < 42.343$ & $ < -4.202$ & $ < 0.756$ & $ < 43.099$ & $ < -3.447$ \\     
  $02$Xund & $0.4500\pm0.0002$ & $2.38$ & $17386.85\pm12139.42^{(\ast)}$ & $39.920\pm0.008^{(\ast)}$ & $9.039\pm0.819^{(\ast)}$ & $ < 42.612$ & $ < -4.541$ & $ < 0.859$ & $ < 43.471$ & $ < -3.682$ \\     
  $03$Xund & $1.0282\pm0.0020$ & $2.93$ & $ 2923.67\pm  254.74$ & $41.857\pm0.005$ & $8.459\pm0.076$ & $ < 43.533$ & $ < -3.040$ & $ < 1.214$ & $ < 44.747$ & $ < -1.826$ \\	
  $04$Xund & $0.9730\pm0.0013$ & $1.89$ & $ 3359.55\pm  766.75$ & $41.547\pm0.011$ & $8.425\pm0.198$ & $ < 44.634$ & $ < -1.905$ & $ < 1.637$ & $ < 46.271$ & $ < -0.268$	
  \\[0.1cm]
 \multicolumn{11}{l}{[\nodata]}
  \\[0.1cm]
\enddata
\tablefoot{
 The complete table is published in its entirety in the electronic edition of the journal.
 A portion is shown here for guidance regarding its form and content.
 In the first part, we list the X-ray detected AGN; in the second part, we report basic data for the $11$ sources in the parent sample undetected in the X-ray (see text).\\
 Col. (1): Unique source identification number in the 2XMMi catalogue; for the X-ray undetected sources, arbitrary ID number (as in Table~\ref{tab:sample}).
 Col. (2): Redshift for the \sdss\ counterpart (as in Table~\ref{tab:sample}).
 Col. (3): Galactic column density, from \citet{nh}; in units of $10^{20}\,$\nh.
 Col. (4): FWHM of the \mgii\ emission line, in units of km~s$^{-1}$; for sources marked with ${(\ast)}$, from the \sdss\ DR7 catalogue, otherwise from \citet{shen10}.
 Col. (5): Monochromatic continuum luminosity at $3000\,$\AA\ rest frame, in units of \lumA; for sources marked with ${(\ast)}$ recovered as described in Sect.~\ref{sect:nucl}, otherwise from \citet{shen10}.
 Col. (6): Mass of the central BH, in units of \pedix{M}{\sun}; for sources marked with ${(\ast)}$, computed from eq.~(\ref{eq:mbh}); otherwise from \citet{shen10}.
 Col. (7): Hard X-ray luminosity, recovered as described in Sect.~\ref{sect:nucl}; in units of \lum\ (upper limit for sources undetected in the X-ray).
 Col. (8): Ratio between the hard X-ray luminosity and the Eddington luminosity, $\pedix{L}{Edd}\equiv1.3\times 10^{38}\,$\pedix{M}{BH}/\pedix{M}{\sun}~[\lum] (upper limit for sources undetected in the X-ray).
 Col. (9): Luminosity-dependent X-ray bolometric correction, from equation~(\ref{eq:lbol}), \citep{marconi04}; upper limit for sources undetected in the X-ray.
 Col. (10): Bolometric luminosity, recovered as described in Sect.~\ref{sect:nucl}; in units of \lum\ (upper limit for sources undetected in the X-ray).
 Col. (11): Eddington ratio, defined as $\lambda\equiv\pedix{L}{bol}/\pedix{L}{Edd}$ (upper limit for sources undetected in the X-ray).}
\end{deluxetable}
\end{landscape}
%


%
\begin{landscape}
\begin{deluxetable}{r c c c@{\extracolsep{-0.001cm}} c@{\extracolsep{-0.001cm}} c@{\extracolsep{-0.001cm}}
c@{\extracolsep{-0.001cm}} c l@{\extracolsep{-0.3cm}} c@{\extracolsep{-0.001cm}} c@{\extracolsep{-0.001cm}} c c}
 \tabletypesize{\tiny}
 \tablecolumns{13}
 \tablewidth{0pt}
 \tablecaption{\footnotesize Radio properties for the $852$ X-ray selected $+11$ X-ray undetected \sdss\ type~1 AGN. \label{tab:radio}}
 \tablehead{
  \colhead{SRCID} & \colhead{\pedix{F}{rf}@$2500\,$\AA} & \colhead{\pedix{\alpha}{ox}} & \colhead{\pedix{F}{rf}@$4400\,$\AA} & \colhead{\pedix{F}{rf}@$5\,$GHz} & \colhead{\pedap{F}{rf}{lim}@$5\,$GHz} & \colhead{$\mathcal{R}$} & \colhead{\apix{\mathcal{R}}{up.lim}} & \colhead{Class.} & \colhead{\pedix{\mathcal{R}}{4400\,\AA}} & \colhead{\pedap{\mathcal{R}}{4400\,\AA}{up.lim}} & \colhead{\pedix{\mathcal{R}}{X}} & \colhead{\pedap{\mathcal{R}}{X}{lim}} \\
  \colhead{(1)} & \colhead{(2)} & \colhead{(3)} & \colhead{(4)} & \colhead{(5)} & \colhead{(6)} & \colhead{(7)} & \colhead{(8)} & \colhead{(9)} & \colhead{(10)} & \colhead{(11)} & \colhead{(12)} & \colhead{(13)} }
\startdata
 \cutinhead{\it X-ray detected AGN}
  $192021$ & $0.0258\pm0.0007$ & $-1.2251\pm0.0292$ & $0.0732\pm0.0018$ &  \emph{undet.}      & $0.6451\pm0.1754$ & \emph{undet.}         & $24.983\pm6.828$ & NC   &  \emph{undet.}   & $ 8.812\pm2.405$ &  \emph{undet.}    & $0.0002\pm0.0001$ \\
  $192103$ & $0.0529\pm0.0015$ & $-1.1479\pm0.0122$ & $ -             $ & $134.9836\pm7.8384$ & $ -             $ & $2550.286\pm164.680 $ & $ -            $ & RL   & $ -	       $ & $ -	        $ & $0.0148\pm0.0013$ & $ -      	    $ \\
  $192150$ & $0.1318\pm0.0037$ & $-1.7358\pm0.044$ & $0.1459\pm0.0037$ & \emph{OutFIRST} & \emph{OutFIRST} & \emph{OutFIRST} & \emph{OutFIRST} & $-$ & \emph{OutFIRST} & \emph{OutFIRST} & \emph{OutFIRST} & \emph{OutFIRST} \\  
  $192335$ & $0.0309\pm0.0009$ & $-1.3782\pm0.2267$ & $0.0616\pm0.0015$ &  \emph{undet.}      & $0.6504\pm0.1721$ & \emph{undet.}         & $21.022\pm5.594$ & NC   &  \emph{undet.}   & $10.556\pm2.806$ &  \emph{undet.}    & $0.0007\pm0.0005$ 
   \\[0.1cm]
 \multicolumn{13}{l}{[\nodata]}
  \\[0.1cm]
  \hline \hline
  \\[0.01cm]
 \cutinhead{\it X-ray undetected AGN}
  $01$Xund & $0.0007\pm0.0001$ & $< -0.9594$ & $0.0281\pm0.0007$ & $0.8359\pm0.1700$ & $ -  $ & $1236.742\pm253.902$ & $ -  $ & RL  & $29.700\pm6.085$ & $ -  $ & $ -  $ & $ > 0.0068$ \\ 
  $02$Xund & $0.0048\pm0.0001$ & $< -1.1943$ & $ -	      $ & $1.6489\pm0.2886$ & $ -  $ & $ 345.411\pm 61.235$ & $ -  $ & RL  & $ -	    $ & $ -  $ & $ -  $ & $ > 0.0104$ \\
  $03$Xund & $0.0712\pm0.0020$ & $< -1.6124$ & $ -	      $ & $0.4992\pm0.1278$ & $ -  $ & $   7.015\pm  1.807$ & $ -  $ & dRI & $ -	    $ & $ -  $ & $ -  $ & $ > 0.0020$ \\
  $04$Xund & $0.0416\pm0.0012$ & $< -1.0559$ & $ -	      $ & $2.8774\pm0.3212$ & $ -  $ & $  69.226\pm  7.971$ & $ -  $ & RL  & $ -	    $ & $ -  $ & $ -  $ & $ > 0.0008$
  \\[0.1cm]
 \multicolumn{13}{l}{[\nodata]}
  \\[0.1cm]
\enddata
\tablefoot{
 The complete table is published in its entirety in the electronic edition of the journal.
 A portion is shown here for guidance regarding its form and content.
 In the first part, we list the X-ray detected AGN; in the second part, we report basic data for the $11$ sources in the parent sample undetected in the X-ray (see text).\\
 Col. (1): Unique source identification number in the 2XMMi catalogue; for the X-ray undetected sources, arbitrary ID number (as in Table~\ref{tab:sample}).
 Col. (2): Monochromatic optical continuum flux at $2500\,$\AA, recovered as described in Sectt.~\ref{sect:nucl} and \ref{sect:rl} (in units of mJy).
 Col. (3): X-ray--to--optical spectral index, $\pedix{\alpha}{ox}\equiv \log (\pedix{F}{2\kev}/\pedix{F}{2500\ang})/\log (\pedix{\nu}{2\kev}/\pedix{\nu}{2500\ang})$; upper limit for sources undetected in the X-ray.
 Col. (4): Monochromatic optical continuum flux at $4400\,$\AA, recovered as described in Sectt.~\ref{sect:nucl} and \ref{sect:sed} (in units of mJy); ``-'': source without \hb\ line in their spectra.
 Col. (5): Monochromatic radio flux at $5\,$GHz (rest frame; in units of mJy), recovered from the integrated flux density at $1.5\,$GHz assuming a power-law spectrum \pedix{F}{$\nu$}$\propto$\apix{\nu}{$-\alpha$} with index $\alpha=0.5$, as described in Sect.~\ref{sect:rl}; \emph{OutFIRST}: source not covered by FIRST; \emph{undet.}: source not detected by FIRST.
 Col. (6): For sources covered but not detected by FIRST, monochromatic radio flux limit at $5\,$GHz (rest frame; in units of mJy), calculated from the flux limit at $1.5\,$GHz of the FIRST survey, $\pedix{F}{1.5\,GHz}=1\,$mJy, assuming the same spectral shape (\pedix{F}{$\nu$}$\propto$\apix{\nu}{$-\alpha$} with index $\alpha=0.5$).
 Col. (7): Radio-UV radioloudness parameter, defined as the ratio between the radio flux at $5\,$GHz and the optical flux at $2500\,$\AA, both rest frame; \emph{OutFIRST}: source not covered by FIRST; \emph{undet.}: source not detected by FIRST.
 Col. (8): For sources covered but not detected by FIRST, limit to the radio-UV radioloudness parameter, calculated from the limit to the radio flux at $5\,$GHz and the optical flux at $2500\,$\AA, both rest frame.
 Col. (9): Radio classification: RL = radioloud; dRI = detected radiointermediate;  ndRI = nondetected with $1 <\mathcal{R}^{up.lim}\leq 10$; RQ = radioquiet; NC = nonclassified; ``-'' = source not covered by FIRST.
 Col. (10): Radio-optical radioloudness parameter, defined as the ratio between the radio flux at $5\,$GHz and the optical flux at $4400\,$\AA, both rest frame; \emph{OutFIRST}: source not covered by FIRST; \emph{undet.}: source with \hb\ line in its spectrum but not detected by FIRST; ``-'': source (detected or not detected by FIRST) without \hb\ line in its spectrum.
 Col. (11): For sources covered but not detected by FIRST, limit to the radio-optical radioloudness parameter, calculated from the limit to the radio flux at $5\,$GHz and the optical flux at $4400\,$\AA, both rest frame; ``-'': source without \hb\ line in its spectrum.
 Col. (12): Radio-X-ray radioloudness parameter, defined as the ratio between the radio luminosity at $5\,$GHz and the X-ray luminosity in the $2-10\,$keV energy range; \emph{OutFIRST}: source not covered by FIRST; \emph{undet.}: source not detected by FIRST.
 Col. (13): For sources covered but not detected by FIRST, limit to the radio-X-ray radioloudness parameter, calculated from the limit to the radio luminosity at $5\,$GHz and the X-ray luminosity in the $2-10\,$keV energy range. Note that for the X-ray detected AGN not detected by FIRST (first part), it is an upper limit, while for sources in the second part, radio-detected and X-ray undetected, it is a lower limit.}
\end{deluxetable}
\end{landscape}
%




\end{document}